\documentclass[11pt]{amsart}
\usepackage{hyperref}
\hypersetup{linktocpage = true, colorlinks = true, linkcolor = blue, citecolor= red, urlcolor = green}

\usepackage{tikz}
\usetikzlibrary{positioning, quotes, arrows, decorations.markings, decorations.pathmorphing, decorations.pathreplacing, shapes, patterns, calc}
\usepackage{url}

\usepackage{graphicx, color, bm}
\usepackage{array,tikz,tikz-cd,float}

\usepackage{amsmath,amsthm}
\usepackage{amssymb, amsfonts, verbatim, subfigure}

\usepackage{mathpazo}
\usepackage{mathrsfs}
\usepackage[mathscr,mathcal]{euscript}

\tikzset{middlearrow/.style={
		decoration={markings,
			mark= at position 0.5 with {\arrow{#1}} ,
		},
		postaction={decorate}
	}
}

\newcommand{\pd}{{\partial}}

\newcommand{\wt}{\widetilde}
\newcommand{\id}{\text{id}}
\newcommand{\Ops}{\text{Ops}}

\newcommand{\be}{\begin{equation}}
	\newcommand{\ee}{\end{equation}}
\newcommand{\beqn}{\begin{equation}}
	\newcommand{\eeqn}{\end{equation}}
\newcommand{\bp}{\begin{pmatrix}}
	\newcommand{\ep}{\end{pmatrix}}
\newcommand{\bsp}{\left(\begin{smallmatrix}}
	\newcommand{\esp}{\end{smallmatrix}\right)}

\newcommand*\diff{\mathop{}\!\mathrm{d}}

\newcommand{\R}{{\mathbb R}}
\renewcommand{\P}{{\mathbb P}}

\newcommand{\C}{{\mathbb C}}
\newcommand{\Z}{{\mathbb Z}}

\newcommand{\CA}{{\mathcal A}}

\newcommand{\CD}{{\mathcal D}}

\newcommand{\CN}{{\mathcal N}}
\newcommand{\CO}{{\mathcal O}}

\newcommand{\CR}{{\mathcal R}}

\newcommand{\CY}{{\mathcal Y}}

\newcommand{\CX}{{\mathcal X}}

\DeclareMathOperator{\Lie}{Lie}
\DeclareMathOperator{\Spin}{Spin}
\newcommand{\fsl}{\mathfrak{sl}}
\newcommand{\fgl}{\mathfrak{gl}}
\newcommand{\fg}{\mathfrak{g}}

\newcommand{\ow}{{\overline{w}}}

\newcommand{\oz}{{\overline{z}}}

\newcommand{\oQ}{{\overline{Q}}}

\newcommand{\bA}{\mathbf{A}}
\newcommand{\bB}{\mathbf{B}}

\newcommand{\bJ}{\mathbf{J}}

\newcommand{\bL}{\mathbf{L}}
\newcommand{\bM}{\mathbf{M}}
\newcommand{\bO}{\mathbf{O}}

\newcommand{\bT}{\mathbf{T}}
\newcommand{\bV}{\mathbf{V}}
\newcommand{\bW}{\mathbf{W}}
\newcommand{\bX}{\mathbf{X}}
\newcommand{\bY}{\mathbf{Y}}
\newcommand{\bZ}{\mathbf{Z}}
\newcommand{\bGamma}{\mathbf{\Gamma}}
\newcommand{\bDelta}{\mathbf{\Delta}}
\newcommand{\bXi}{\mathbf{\Xi}}
\newcommand{\bSigma}{\mathbf{\Sigma}}
\newcommand{\bTheta}{\mathbf{\Theta}}
\newcommand{\bPsi}{\mathbf{\Psi}}
\newcommand{\bPhi}{\mathbf{\Phi}}

\newcommand{\bLambda}{\mathbf{\Lambda}}
\newcommand{\bmu}{\boldsymbol{\mu}}
\newcommand{\bnu}{\boldsymbol{\nu}}
\newcommand{\bOmega}{\mathbf{\Omega}}

\numberwithin{equation}{section}
\numberwithin{figure}{section}
\numberwithin{table}{section}

\newcommand{\lie}{\mathfrak}
\newcommand{\cN}{\mathcal{N}}
\newcommand{\til}{\wt}
\newcommand{\bu}{\bullet}
\newcommand{\del}{\partial}

\newcommand{\op}{\operatorname}
\newcommand{\cO}{\CO}

\newcommand{\zbar}{\Bar{z}}
\renewcommand{\d}{\text{d}}
\newcommand{\xto}{\xrightarrow}
\newcommand{\cA}{\mathcal{A}}
\newcommand{\define}{\overset{\text{def}}{=}}

\DeclareMathOperator{\Sym}{Sym}

\DeclareMathOperator{\Tr}{Tr}
\DeclareMathOperator{\Vect}{\text{Vect}}

\DeclareMathOperator{\htl}{\{\!\{}
\DeclareMathOperator{\htr}{\}\!\}}

\usepackage{todonotes}

\newtheoremstyle{thm}
{7pt}
{7pt}
{\itshape}
{}
{\bf}
{.}
{5pt}
{\thmnumber{#2 }\thmname{#1}\thmnote{ (#3)}}

\newtheoremstyle{def}
{7pt}
{10pt}
{\itshape}
{}
{\bf}
{.}
{5pt}
{\thmnumber{#2} \thmname{#1}\thmnote{ (#3)}}

\newtheoremstyle{rem}
{4pt}
{10pt}
{}
{}
{\itshape}
{:}
{3pt}
{}

\newtheoremstyle{texttheorem}
{8pt}
{8pt}
{\itshape}
{}
{\bf}
{. \hspace{5pt}}
{3pt}
{}

\theoremstyle{thm}

\newtheorem*{theorem*}{Theorem}
\newtheorem*{lemma*}{Lemma}
\newtheorem*{corollary*}{Corollary}
\newtheorem*{proposition*}{Proposition}
\newtheorem*{definition*}{Definition}

\newtheorem*{conjecture}{Conjecture}

\newtheorem{theorem}{Theorem}[subsection]
\newtheorem{thm-def}{Theorem/Definition}[theorem]
\newtheorem{proposition}[theorem]{Proposition}

\newtheorem*{question*}{Question}

\numberwithin{equation}{subsection}

\theoremstyle{def}

\theoremstyle{rem}

\usepackage{stmaryrd}
\date{}

\title{Enhanced symmetries in minimally twisted three-dimensional supersymmetric theories}
\author{Niklas Garner}
\address{University of Washington, Seattle}
\email{nkgarner@uw.edu}

\author{Surya Raghavendran}
\address{Yale University}
\email{surya.raghavendran@yale.edu}

\author{Brian R. Williams}
\address{Boston University}
\email{bwill22@bu.edu}

\begin{document}
	
\begin{abstract}
We show that the action of residual supersymmetries in holomorphic-topological twists of $\cN=2$ theories in three dimensions naturally extends to the action of certain infinite dimensional Lie superalgebras. We demonstrate this in a range of examples, including $\cN=4$ Yang-Mills theories and superconformal Chern-Simons theories, describing how the symmetries are implemented at the level of local operators.
\end{abstract}

\maketitle
\tableofcontents

\section{Introduction}
The utility of symmetry is a recurring theme in the study of quantum field theories. In many situations where exact computations are possible, their existence is owed to the presence of a nonmanifest infinite dimensional symmetry - the enhancement of conformal symmetries in two-dimensional conformal field theory is perhaps the most well-known illustration of this paradigm. In this paper, we show that supersymmetric field theories in three-dimensions admit such a symmetry enhancement, after performing a \textit{holomorphic-topological twist}.

Twisting refers to a localization or fixed-point procedure for supersymmetric field theories, whereby one considers the invariants of a theory with respect to a square-zero supercharge $Q$. Operationally, this is accomplished by modifying the BRST differential of the theory by $Q$. Local operators in the cohomology of the modified BRST differential are some fractionally BPS operators and will be such that their correlation functions are killed by those infinitessimal translations in the image of $Q$. More invariantly, the image of bracketing by $Q$ defines a transversely-holomorphic-foliation (THF) and the twist is a simpler theory where all correlation functions are constant along the leaves. 

Whenever a theory admits a twist, it admits a minimal, or holomorphic-topological ($HT$) twist, where the rank of the THF is minimal, and a maximal amount of information about the original theory is retained. More commonly studied topological twists arise as further deformations of an $HT$ twist. In three-dimensions, $HT$ twists exist given $\cN=2$ supersymmetry, and the result is a theory that can be placed on THF 3-manifolds locally modeled on $\C \times \R$. 

A twist of any supersymmetric field theory will always retain an action of the commutant of the supercharge used to twist. We show in three-dimensional examples that the action of this commutant, a priori a finite dimensional subalgebra of the supersymmetry algebra, extends to the action of an infinite dimensional algebra. An analogous enhancement was studied in \cite{SWsuco} for the minimal twist of four-dimensional supersymmetric field theories.

Explicitly, in Section \ref{sec:N=2} we show that for a general class of three-dimensional $\cN=2$ theories that upon performing the $HT$ twist the resulting theory locally on $\C \times \R$ shares the same symmetries of a chiral CFT on $\C$, namely holomorphic vector fields along $\C$.\footnote{On a more general THF, one should consider vector fields which are constant along the leaves; in this simple split case this is precisely the holomorphic vector fields on $\C$.}
Global symmetries present in the three-dimensional supersymmetric theory receive a similar enhancement. For example, if a three-dimensional theory is equipped with a flavor symmetry by some group $H$ then locally on $\C \times \R$ the $HT$ twist is equipped with an infinitesimal symmetry by the Lie algebra of holomorphic functions on $\C$ with values in $\lie{h}$.

In conformal field theory, the symmetry by the algebra of holomorphic vector fields is further extended to an action of the Virasoro algebra, which is a central extension of holomorphic vector fields on the punctured plane $\C^\times$. In the $HT$ twist of a three-dimensional supersymmetric theory there is a similar phenomenon. Instead of holomorphic vector fields on $\C^\times$, this further enhanced algebra is an extension of the (derived) vector fields on punctured three-dimensional space $\C \times \R - \{0\}$ which are constant along the leaves of the standard THF. We briefly elaborate on this phenomenon in Section~\ref{sec:descent}. Following the analogy with conformal field theory, we point out that the method of descent leads to a three-dimensional version of a vertex algebra which the first and third author have developed mathematically in \cite{GWrav}. We will further pursue this point of view in more supersymmetric examples in a companion paper \cite{GRW2}.

Any theory with $\cN \geq 2$ supersymmetry can be thought of as a theory with $\cN=2$ supersymmetry so that its $HT$ twist may be considered and its enhanced symmetries examined. We begin doing so in Section \ref{sec:SYM}, with the example of $\cN=4$ Yang-Mills theories, and we find that the above enhancement to holomorphic vector fields on $\C$ gets enhanced yet further to the positive part of the $\cN=2$ Virasoro algebra. We then turn to understanding further deformations of $\cN=4$ theories from the point of view of the enhanced symmetry algebras present in the $HT$ twist. This includes the $A$ and $B$ twists, studied extensively in the context of three-dimensional mirror symmetry, as well as $\Omega$-background deformations. 

However, as the odd part of the enhanced symmetry algebra is infinite dimensional, there are many more deformations that can be considered. We highlight one family of such deformations, which generalize the $\Omega$-backgrounds in that they localize the $HT$ twisted theory to a $n$-th order neighborhood of $\{0\} \times \subset \C \times \R$. Finally, we discuss how an involution of the supersymmetry algebra that exchanges the $A$ and $B$ twists extends to the entire infinite-dimensional enhanced symmetry algebra. This involution can be viewed as part of the basic parameter-matching of three-dimensional mirror symmetry, and its extension suggests interesting generalizations.

In Section \ref{sec:CSM} we turn our attention to superconformal Chern-Simons-matter theories, beginning with the $\cN=3$ theories found separately by Zupnik-Khetselius and Kao-Lee  \cite{ZupnikKhetselius, KaoLee}; see also \cite{GaiottoYin, KSmirrorsym}. We find that the $HT$ twist of such theories admits an action of the positive modes of the $\cN=1$ Virasoro algebra. We then progressively constrain these theories to increase the amount of supersymmetry, cf. \cite{dMFOME}: there are the $\CN=4$ theories of Gaiotto-Witten \cite{GaiottoWitten-Janus} as well as their generalizations including twisted hypermultiplets due to Hosomichi-Lee-Lee-Lee-Park \cite{HLLLP1}, where we again find the positive modes of the $\CN=2$ Virasoro algebra. The $\CN=5$ and $6$ theories of Hosomichi-Lee-Lee-Lee-Park \cite{HLLLP2} come next, and we find the positive modes of the $\CN=3$ and big $\CN=4$ Virasoro algebras, respectively.

Most strikingly, we find that in theories with $\cN=8$ supersymmetry, the $HT$ twist has a symmetry by the exceptional Lie superalgebra $E(1|6)$ which appears in Kac's classification of infinite dimensional linearly compact Lie superalgebras \cite{KacClass}. 
This same exceptional Lie superalgebra was found by the last two authors as an asymptotic symmetry of a twisted version of the $AdS_4\times S^7$ background of eleven-dimensional supergravity~\cite{twistedgraviton}. 
We discuss two examples of an enhancement by $E(1|6)$: the BLG theory \cite{BL1, BL2, Gustavsson} and the rank 1 ABJM theory at levels $k=1,2$ \cite{ABJM, ABJ}. While in the former example, the presence of $\cN=8$ supersymmetry is visible from the Lagrangian, the latter example has $\cN=8$ supersymmetry due to a nonperturbative effect, and accordingly, currents realizing the enhanced symmetry involve monopoles. 

This last series of examples serves as an illustration of how the $HT$ twist can be used to diagnose supersymmetry enhancements. Namely, if the expected action of holomorphic vector fields in the $HT$ twist is extended to (e.g.) the positive part of the $\CN=2$ Virasoro algebra then this is a good indication that the theory has (or flows to a theory with) $\CN=4$ supersymmetry. We note that this enhancement is necessary but not sufficient: the $HT$ twist of a free $\CN=2$ chiral multiplet (or multiplet copies thereof) has this symmetry, cf. Section 2.4.4 of \cite{GRW2}, but a free chiral certainly doesn't have $\CN=4$ supersymmetry.

We point out that while our arguments for enhanced symmetries of $HT$ twists of theories with $\cN\geq 2$ symmetry are classical, the statements can be interpreted at the quantum level (at least perturbatively).
Indeed, one of the main results of \cite{GRWthf} is that symmetries of holomorphic-topological theories on $\C^n \times \R^m$, $m \geq 1$ are non-anomalous to one-loop in perturbation theory.
Since the $HT$ twist of the supersymmetric theories we study are all exact at one-loop, we can apply this result to the case $n=m=1$ to obtain the quantum version of the statements in this paper.

A summary of enhanced symmetry algebras present in the $HT$ twist is given in Table \ref{tab:enhanced}.
In this table we choose to highlight how the twist of the superconformal algebra is enhanced.

\begin{table}
\centering
\begin{tabular}{ |c|c|c|c|c|c}
 \hline
 $\cN=k$ & $\lie{sconf}_{\cN=k}$ & $(\lie{sconf}_{\cN=k})^{Q_{HT}}$ & $\lie{a}_{\cN=k}$ & $\lie{f}_{\cN=k}$ \\
 \hline
 $\cN=2$   & $\lie{osp}(2|4)$   & $\lie{sl}(2)$ & $\lie{vir}^{\geq 0}$ & $\lie{h} \otimes \cO(\C)$ \\
 $\cN=3$ &   $\lie{osp}(3|4)$  & $\lie{osp}(1|2)$   & $\lie{vir}_{\cN=1}^{\geq 0}$ & $-$ \\
 $\cN=4$ &   $\lie{osp}(4|4)$  & $\lie{osp}(2|2)$   & $\lie{vir}_{\cN=2}^{\geq 0}$ & $\lie{h} \otimes \cO(\C^{1|1})$ \\
 $\cN=5$ &   $\lie{osp}(5|4)$  & $\lie{osp}(3|2)$   & $\lie{vir}_{\cN=3}^{\geq 0}$ & $-$ \\
 $\cN=6$ & $\lie{osp}(6|4)$ & $\lie{osp}(4|2)$ &  $K'_4{}^{\geq 0}$ & $-$ \\
 $\cN=8$    &$\lie{osp}(8|4)$ & $\lie{osp}(6|2)$ & $E(1|6) = CK_6^{\geq 0}$ & $-$ \\
 \hline
\end{tabular}
\caption{Symmetry enhancement for HT twists of three-dimensional $\cN=k$ theories on $\C \times \R$.
The right hand column denotes the enhanced flavor symmetry algebra of a three-dimensional $\cN=k$ theory with flavor symmetry group $H$ and Lie algebra $\lie{h}$.}
\label{tab:enhanced}
\end{table}

\subsection*{Acknowledgements}
We would like to thank Kevin Costello, Tudor Dimofte, Zhengping Gui, Natalie M. Paquette, Ingmar Saberi, and Keyou Zeng for useful conversations during the development of these ideas. 
NG is supported by funds from the Department of Physics and the College of Arts \& Sciences at the University of Washington, Seattle.
SR was supported by Yale University and the Perimeter Institute for Theoretical Physics. Research at Perimeter Institute is supported in part by the Government of Canada, through the Department of Innovation, Science and Economic Development Canada, and by the Province of Ontario, through the Ministry of Colleges and Universities.

\section{Twisted Three-dimensional Supersymmetry}
\label{sec:N=2}
In this section we review and introduce the essential ingredients of our approach to deforming the holomorphic-topological ($HT$) twist of three-dimensional $\CN = 2$ theories to topological theories following \cite{AganagicCostelloMcNamaraVafa,CostelloDimofteGaiotto-boundary}.

\subsection{Twisted formalism for three-dimensional $\CN = 2$ theories}
\label{sec:three-dimensionaltwisted}
We begin by discussing the essential features of the twisted formalism of loc. cit. 
The utility of this twisted formalism is to dramatically simplify the field content of the theory without losing any of the derived structures admitted by local and extended operators, e.g., higher operations obtained by descent.

\subsubsection{Twisting}
The starting point for us is three-dimensional supersymmetry.
In Euclidean signature, the complexified three-dimensional $\CN=k$ supersymmetry algebra is the Lie superalgebra whose underlying super vector space is
\be
\C^3 \oplus \Pi S \otimes W
\ee
where $S$ is the two-dimensional irreducible spin representation for $\lie{so}(3;\C)$ and $W$ is a $k$-dimensional vector space equipped with a bilinear form.
The nontrivial Lie bracket is determined by the unique $\lie{so}(3;\C)$ invariant bilinear form $S \times S \to \C$ together with the bilinear form on $W$.

In this section we will be concerned with the case of $\cN=2$ supersymmetry.
In this case, the algebra admits an $R$-symmetry by the group $U(1)_R$ which acts on $W$ with weights $\pm 1$ with respect to a chosen orthogonal basis.
In such a basis, the algebra has four odd generators $Q_\alpha, \oQ_\alpha$, $\alpha = \pm$, with super bracket
\be
[Q_\alpha, \oQ_\beta] = (\sigma^\mu)_{\alpha \beta} P_{\mu}\,,
\ee
where $(\sigma^\mu)^\alpha{}_\beta$ are the Pauli matrices and $P_\mu, \mu=1,2,3$ generate infinitesimal translations of three-dimensional Euclidean space.\footnote{Spinor indices are raised and lowered using the Levi-Civita symbol as $\chi_\alpha = \epsilon_{\alpha \beta} \chi^\beta$ and $\chi^\beta = \chi_\alpha \epsilon^{\alpha \beta}$, where $\epsilon_{+-} = \epsilon^{+-} = 1$.} 
The $R$-symmetry assigns $Q_\alpha, \oQ_\alpha$ weights $-1, 1$ respectively. 

Up to spacetime symmetries and $R$-symmetry rotations, there is a unique nilpotent supercharge in this algebra and hence twist \cite{EagerSaberiWalcher, EStwists}, which we take to be 
\be
Q_{HT} = \oQ_+ .
\ee 
If we choose coordinates on Euclidean space $\R^3 = \C_{z,\oz} \times \R_t$ as
\be
	z = x_1+i x_2, t = x_3,
\ee
the non-trivial brackets involving $Q_{HT}$ are given by
\be
[Q_{HT}, Q_\oz] = P_{\oz} \qquad [Q_{HT}, Q_t] = P_t\,,
\ee
where $Q_\oz = \tfrac{1}{2} Q_+$ and $Q_t = -Q_-$. 
Thus, the cohomology with respect to $Q_{HT}$ will behave holomorphically on $(z,\zbar)$ and topologically in the coordinate $t$.
For this reason, a twist by this element is called a ``holomorphic-topological" ($HT$) twist. It is sometimes called the ``holomorphic twist,'' cf. \cite{CostelloDimofteGaiotto-boundary}.

The $Q_{HT}$ twist is compatible on three-dimensional manifolds that locally admit a coordinatization of the form $\C_{z,\oz} \times \R_t$ (or $\C_{z,\oz} \times \R_{t \geq 0}$). More precisely, the $\CN=2$ theories we will be interested in preserve the full $U(1)_R$ $R$-symmetry. 
With only a $U(1)_R$ $R$-symmetry, it is not possible to define the $Q_{HT}$ twist on an arbitrary three-manifold. 
Instead, we can work on a three-manifold compatible with reduction of the Lorentz group to the subgroup $\Spin(2)_E \subset SU(2)_E$ preserving vectors tangent to $\C$. 
The interiors of such manifolds locally take the form $\C_{z,\oz} \times \R_t$ and transition functions between patches $\C_{z,\oz} \times \R_t$ and $\C_{z',\oz'} \times \R_{t'}$ are of the form
\be
z \to z'(z) \qquad \oz \to \oz'(\oz) \qquad t \to t'(z,\oz,t).
\ee
Boundaries are identical, but instead are modeled on $\C_{z,\oz} \times \R_{t\geq 0}$. Such data equips the three-manifold with a transverse holomorphic foliation (THF).

With respect to the subgroup $\Spin(2)_E$, the supercharges $Q_\pm, \oQ_\pm$ have spin $J_0 = \pm \tfrac{1}{2}$. The twisting homomorphism, simply amounts to working with respect to the ``twisted spin" $\Spin(2)_{E'}$ generated by $J$ given by
\be
J = \tfrac{1}{2} R - J_0.
\ee
With this choice, the supercharge $Q_{HT}$ has twisted spin $J=0$ and $U(1)_R$ $R$-charge $R = 1$. Similarly, the supercharges $Q_\oz$ and $Q_t$ have $U(1)_R$ $R$-charge $R = -1$ and twisted spins $J=-1$ and $J=0$, respectively.

\subsubsection{BV-BRST and twisted superfields}
In the following sections we will be interested in $\CN=2$ theories of vector multiplets  coupled to matter fields transforming as chiral superfields. 
We will assume that these theories are equipped with an $R$-symmetry by the group $U(1)$.
See, e.g., \cite{AHISS} for a review of the untwisted $\CN=2$ theories. Here, we focus on the description of $HT$-twisted $\CN=2$ Chern-Simons--matter theories in \cite{CostelloDimofteGaiotto-boundary,AganagicCostelloMcNamaraVafa}, which uses the Batalin-Vilkovisky (BV) formalism~\cite{BV,CostelloRenormalization}.

The data of a three-dimensional $\CN=2$ Chern-Simons--matter theory is the following:
\begin{itemize}
	\item A compact gauge group $G_\R$.
	\item A unitary representation $V$ of $G_\R$, a $G_\R$-invariant superpotential $W \colon V \to \C$ of $R$-charge $2$, and a collection of Chern-Simons levels $k$. 
\end{itemize}

We assume that the matter representation $V$ decomposes as $V = \bigoplus_r V^{(r)}$, where the $R$-symmetry $U(1)_R$ acts by weight $r$ on $V^{(r)}$. 
Denote by $\fg_\R = \Lie(G_\R)$ the (real) Lie algebra of $G_\R$, $G$ the complexification of $G_\R$, and $\fg = \Lie(G)$ its (complex) Lie algebra.

Using the notation of \cite{CostelloDimofteGaiotto-boundary}, we define the graded vector space
\beqn
\bOmega^{(j),\bullet} := C^\infty(\R^3)[\diff t, \diff \oz] \diff z^j .
\eeqn
We view $\d t, \diff \oz$ as elements of cohomological degree $+1$ and $\d z^j$ is cohomological degree zero.
Although we work locally for now, there is a similar complex defined for any three-manifold equipped with a THF, see section \ref{sec:descent}. 
There is a natural (wedge) product $\wedge \colon \bOmega^{(j),i} \otimes \bOmega^{(j'),i'} \to \bOmega^{(j+j'),i+i'}$; in particular when $j=0$ this endows $\Omega^{(0),\bu}$ with the structure of a graded commutative algebra. 
We also utilize an integration map $\int \colon \bOmega^{(1),2}_c  \to \C$, where the subscript denotes forms with compact support. 
There is natural differential operator
\beqn
\diff' = \pd_t \diff t + \pd_{\oz} \diff \oz \colon \Omega^{(j),\bu} \to \Omega^{(j),\bu+1}
\eeqn
of cohomological degree 1, square-zero, and of (twisted) spin $J=0$.
When $j = 0$ this differential endows $\Omega^{(0),\bu}$ with the structure of a commutative dg algebra.
We will also utilize the differential operator
\beqn
\del = \d z \frac{\del}{\del z} \colon \Omega^{(j),\bu} \to \Omega^{(j+1),\bu}
\eeqn
which is of cohomological degree 0 and (twisted) spin $J=1$.

The twisted formalism of this class of theories includes the following fields $Q_{HT}$-closed fields:
\begin{itemize}
	\item two components of the gauge field organized into the fermionic field $$A = A_t \diff t + A_\oz \diff \oz \in \bOmega^{(0),1} \otimes \mathfrak{g},$$ with $A_t$ complexified by the real scalar $\sigma$ of the $\CN=2$ vector multiplet.
	\item a coadjoint-valued bosonic field $$B = B_z \diff z \in \bOmega^{(1),0} \otimes \mathfrak{g}^*,$$ identified in the physical theory with the curvature $\tfrac{1}{g^2} F_{zt}$ up to Chern-Simons terms.
	\item a $V$-valued bosonic field $$\phi = \sum\limits_r \phi_{r} \diff z^{r/2} \in \bigoplus\limits_{r} \bOmega^{(r/2),0} \otimes V^{(r)},$$ identified with the bosons in the chiral superfields after applying the twisting homomorphism turning $\Spin(2)_E$ scalars of $R$-charge $R = r$ to sections of $K_\C^{r/2}$.
	\item a $V^*$-valued one-form valued field $$\eta = \sum\limits_r \big(\eta_{r,t} \diff t + \eta_{r,\oz} \diff \oz \big) \diff z^{1-r/2} \in \bigoplus\limits_{r} \bOmega^{(1-r/2),1} \otimes (V^{(r)})^*,$$ whose components are identified with the covariant derivatives of the conjugate scalar $\eta_t \propto \overline{D_\oz \phi}, \eta_\oz \propto \overline{D_t \phi}$.
\end{itemize}
The components of $A$ and $B$ have $R$-charge $0$, $\phi_r$ has $R$-charge $r$, and $\eta_r$ has $R$-charge $-r$. In the BV formalism, we further include anti-fields $A^*, B^*, \phi^*, \eta^*$ for the fields $A, B, \phi, \eta$ and a differential $Q_{BV}$ schematically given by
\be
Q_{BV}(\text{anti-field}) = \text{EOM for field} \qquad Q_{BV}(\text{field}) = \text{EOM for anti-field}\,.
\ee
The action for our twisted theory takes the form
\be
\label{eq:twistedS}
S = \int B F'(A) + \eta \diff'_A \phi + \tfrac{1}{2} (\eta^*)^2 \pd^2 W + \tfrac{k}{4\pi}\Tr(A \pd A),
\ee
where $\diff'_A = \diff' + A = (\pd_t + A_t) \diff t + (\pd_{\oz} + A_\oz) \diff \oz$ is the covariant derivative, $F'(A) = \diff'A + A^2$ is the corresponding curvature. An even better description of this variation uses the (shifted-)Poisson bracket on the space of fields called the $BV$-bracket $\{-,-\}_{BV}$ and pairs fields and anti-fields as
\be
\{\text{field}, \text{anti-field}\}_{BV} = \delta^{(3)} \diff {\rm Vol}\,   
\ee
from which one identifies $Q_{BV} = \{-,S\}_{BV}$. It is straight-forward to derive the action of $Q_{BV}$ from either of these descriptions. For example, the fields transform as
\be
\begin{aligned}
	Q_{BV} A & = 0 & \qquad Q_{BV} B & = 0\\
	Q_{BV} \phi & = 0 & \qquad Q_{BV} \eta & = \eta^* \pd^2 W\\
\end{aligned}
\ee
because the only anti-field that appears explicitly in the action is $\eta^*$.

This action has two types of redundancies, which the BV formalism accounts for by including ghost fields (and their corresponding anti-fields, aka anti-ghosts). The first is a familiar gauge redundancy, for which we introduce the usual BRST ghost $c$ (a $\mathfrak{g}$-valued, fermionic scalar with $R$-charge $R = 0$: $c \in \bOmega^{(0),0} \otimes \mathfrak{g}$), under which the fields transform as
\be
\begin{aligned}
	\delta_c A & = \diff'_A c & \qquad  \delta_c B & = c\cdot B + \tfrac{k}{2\pi} \pd c\\
	\delta_c \phi & = c\cdot \phi & \qquad \delta_c \eta & = c\cdot \eta
\end{aligned}\,,
\ee
where $c \, \cdot$ denotes the infinitesimal action of $\mathfrak{g}$ with parameter $c$. The unusual variation of $B$ ensures that (for nonzero level $k$) $\tfrac{2\pi}{k} B_z, A_\oz, A_t$ transform as components of a full gauge field $\CA$ that transforms as $\delta_c \CA = \diff_\CA c$, cf. \cite{AganagicCostelloMcNamaraVafa}.

The second redundancy comes from the twisting supercharge itself. In particular, we introduce a $M^*$-valued, fermionic ghost $\psi = \sum_r \psi_r \diff z^{1-r/2} \in \bigoplus_r \bOmega^{(1-r/2),0} \otimes (M^{(r)})^*$ and transformations
\be
\begin{aligned}
	\delta_\psi A & = 0 & \qquad \delta_\psi B & = -\mu\\
	\delta_\psi \phi & = 0 & \qquad \delta_\psi \eta & = \diff'_A \psi + \pd^2 W \eta
\end{aligned}\,,
\ee
where $\mu$ is the moment map for the $\mathfrak{g}$ action on the representation $T^*[1] M \cong M^*[1] \oplus M$; in components it reads $\mu_a = \psi_m (\tau_a)^m{}_n \phi^n$. In the physical theory, $\psi$ can be identified with one of the fermions in the chiral multiplets, after applying the twisting homomorphism.

There is a ghost number symmetry $U(1)_{\rm gh}$, under which the fields $A$, $B$, $\phi$, $\eta$ have ghost number ${\rm gh} = 0$, the anti-fields $A^*$, $B^*$, $\phi^*$, $\eta^*$ have ghost number ${\rm gh} = -1$, the ghosts $c, \psi$ have ghost number ${\rm gh} = 1$, and the anti-ghosts $c^*$, $\psi^*$ have ghost number ${\rm gh} = -2$. We define the cohomological grading $U(1)_C$ as the sum of $R$-charge and ghost number:
\be
C = R + {\rm gh}.
\ee
It is also important to note that we are free to redefine the cohomological grading $C$ by mixing with other abelian symmetries of the theory, we will make use of this freedom below. The twisted theory is thus graded by parity (fermionic or bosonic), twisted spin (generated by $J$), and the cohomological grading (generated by $C$).\footnote{We work in conventions such that parity alone determines the graded-commutativity of observables. Indeed, the $R$-charge, and hence the cohomological grading, in the above class of $\CN=2$ theories may be non-integral.} Each of the variations $\delta_c$ and $\delta_{\psi}$ is fermionic and has cohomological grading $C = 1$ and twisted spin $J = 0$, as desired. We denote the total transformation by
\be
	Q = Q_{BV} + \delta_c + \delta_\psi\,.
\ee

After introducing anti-fields, ghosts, and anti-ghosts, the above field theory can be concisely repackaged in terms of ``twisted superfields." Consider the transformations of $c, A$, and $B^*$; they are given by
\be
Q c = c^2 \qquad Q A = \diff'_A c \qquad Q B^* = c \cdot B^* + F'(A)\,.
\ee
If we define $\bA = c + A + B^* \in \bOmega^{\bullet, (0)} \otimes \fg[1]$, where $[1]$ denotes a shift in cohomological degree by 1, these variations can be neatly repackaged as
\be
Q \bA = F'(\bA) = \overset{\text{0-form}}{c^2} + \overset{\text{1-form}}{\diff'_A c} + \overset{\text{2-form}}{c \cdot B^* + F'(A)}.
\ee
We can similarly combine the remaining fields:
\be
\begin{aligned}
	\bB & = B + A^* + c^* \in \bOmega^{(1),\bu} \otimes \fg^*\\
	\bPhi_r & = \phi_r + \eta_r^* + \psi_r^* \in \bOmega^{(r/2),\bu} \otimes V^{(r)}\\
	\bPsi_r & = \psi_r + \eta_r + \phi_r^* \in \bOmega^{(1-r/2),\bu} \otimes (V^{(r)})^*[1]
\end{aligned}
\ee
The variation of these twisted superfields under $Q_{HT}$ will be given below in Eq. \eqref{eq:QHT}. The twisted spin $J$ and cohomological grading $C$ of these twisted superfields, as well as $\diff t, \diff \oz, \diff z,$ are collected in Table \ref{table:spincoho}.
\begin{table}[h!]
	\centering
	\begin{tabular}{c|c|c|c|c|c|c|c}
		& $\bA$   & $\bB$  & $\bPhi_r$      & $\bPsi_r$          & $\diff t$ & $\diff \oz$ & $\diff z$ \\ \hline
		$(J,C)$ & $(0,1)$ & $(1,0)$ & $(\tfrac{r}{2}, r)$ & $(1-\tfrac{r}{2}, 1-r)$ & $(0,1)$   & $(-1,1)$    & $(1,0)$ 
	\end{tabular}
	\caption{Twisted spin $J$ and cohomological grading $C$ of the twisted superfields and differential forms $\diff t, \diff \oz, \diff z$ in the holomorphic-topological twist. The twisted superfields $\bA,\bB$ come from an $\CN=2$ vector multiplet and $\bPhi, \bPsi$ come from an $\CN=2$ chiral multiplet with $R$-charge $r$.}
	\label{table:spincoho}
\end{table}

As mentioned above, the desired variations arise from the BV-bracket $\{\,-,-\,\}_{BV}$, which pairs fields and anti-fields, via $Q = \{-, S\}_{BV}$. In terms of the twisted superfields, the BV-bracket is explicitly given by
\be
\label{eq:BVbracket}
\{\bA(x),\bB(y)\}_{BV} = \{\bPhi(x), \bPsi(y)\}_{BV} = \delta^{(3)}(x-y) \diff {\rm Vol}\,.
\ee
The appropriate action can be neatly expressed in terms of the above twisted superfields as
\be
\label{eq:twistedaction}
S = \int \bB F'(\bA) + \bPsi \diff'_{\bA} \bPhi + \bW + \tfrac{k}{4 \pi} \Tr(\bA \pd \bA),
\ee
where $\bW = W(\bPhi)$. The $Q_{HT}$ variation of the twisted superfields is then given by
\be
\begin{aligned}
	\label{eq:QHT}
	Q \bA & = \frac{\delta S}{\delta \bB} =  F'(\bA) \qquad & Q \bB & = \frac{\delta S}{\delta \bA} =  \diff'_{\bA} \bB - \bmu + \tfrac{k}{2\pi} \pd \bA\\
	Q \bPhi & = \frac{\delta S}{\delta \bPsi} = \diff'_{\bA} \bPhi \qquad & Q \bPsi & = \frac{\delta S}{\delta \bPhi} = \diff'_{\bA} \bPsi + \frac{\pd \bW}{\pd \bPhi}
\end{aligned}\,,
\ee
where $\bmu = \mu(\bPhi, \bPsi)$.

Nilpotence of $Q$ and invariance of $S$ under $Q$ are equivalent to $S$ solving the classical master equation \cite{CostelloDimofteGaiotto-boundary}. Moreover, $Q^2 = 0$ {\em off-shell} by construction. The theories described by this twisting procedure are a (``chiral") deformation of the class of theories studied in \cite{GwilliamWilliams}, and their results show that these theories have consistent quantization, i.e., it satisfies the quantum master equation (at any scale).%
\footnote{The papers \cite{GwilliamWilliams, GRWthf} use the machinery of the homotopy RG flow of \cite{CostelloRenormalization} to make mathematically precise statements in perturbative theory. In this paper, we will almost entirely ignore these details and work at infinitely long length scales $L \to \infty$ to simplify the discussion.}

\subsection{Spacetime and flavor symmetry enhancement}
\label{sec:currents}

The twisted superfields of \cite{CostelloDimofteGaiotto-boundary,AganagicCostelloMcNamaraVafa} help make holomorphic-topological descent manifest; if $\CO$ is the lowest component of a twisted superfield $\bO$ with $Q \bO = \diff' \bO$, then $\CO$ is $Q$-closed and the higher form components of $\bO$ describe the operators making $\pd_\oz \CO, \pd_t \CO$ cohomologically trivial. On the other hand, for a general $HT$-twisted theory, the operator $\pd_z \CO$ must be realized as a surface integral against the stress tensor:
\be
	\pd_z \CO(z,\oz, t) = \oint_{S^2} *(T_{z \mu} \diff x^\mu) \CO(z,\oz,t)
\ee
In the twisted formalism, the stress tensor $*(T_{z \mu} \diff x^\mu)$ can be expressed as the 2-form component of the twisted superfield $\bT$, although one often needs to modify the na{\"i}ve stress tensor obtained via a Noether procedure on the action in Eq. \eqref{eq:twistedaction}. This (modified) stress tensor is such that $Q \bT$ is $\diff'$-exact, i.e.
\be
	Q \bT = \diff' \bT\,, 
\ee
and hence the surface integral realization of $\pd_z$ is therefore $Q$-closed due to Stokes' theorem \cite[Sec 2.2]{CostelloDimofteGaiotto-boundary}.

The (modified) stress tensor $\bT$ is actually the first of an infinite tower of conserved currents that generates an action of holomorphic vector fields on the space of fields of our $HT$-twisted $\CN=2$ theories. Explicitly, the modified stress tensor $\bT$ is given by
\be
\bT = \iota_{\pd_z} \bigg(-\bB \pd \bA + \sum\limits_r (1-\tfrac{r}{2})\bPsi_r \pd \bPhi_r - \tfrac{r}{2} \bPhi_r \pd \bPsi_r\bigg)\,,
\ee
where $\iota_{\pd_z}$ denotes contraction with $\pd_z$. Verifying $Q \bT = \diff \bT$ uses the fact that the superpotential has $R$-charge 2.

It is important to note that $BV$-bracket with $\int\!\bT$ acts on all fields as $\pd_z$:
\be
	\{\textstyle{\int}\bT, -\}_{BV} =  \pd_z\,.
\ee
Equivalently, the integrated local functional $\int\!\bT$ is the Hamiltionian/moment map for the action of holomorphic translations. This is the (classical) BV realization to the above (quantum) equation -- the stress tensor $\bT$ is a conserved current that generates the action of holomorphic translations. Of course, we can replace $\pd_z$ by a more general holomorphic vector field $V = V(z) \pd_z$ to get yet more currents 
\beqn
\bT_V = \iota_{V} \bigg(-\bB \pd \bA + \sum\limits_r (1-\tfrac{r}{2})\bPsi_r \pd \bPhi_r - \tfrac{r}{2} \bPhi_r \pd \bPsi_r\bigg)\,,
\eeqn
that serve as conserved currents realizing an action of holomorphic vector fields on $\C$, denoted $\textrm{Vect}(\C)$. A straightforward computation verifies that the action of $V$ is encoded by the Lie derivative of forms along $V$:
\be
\begin{aligned}
	V \cdot \bA & = \{\textstyle{\int} \bT_V, \bA\}_{BV} = V \pd_z \bA\\
	V \cdot \bB & = \{\textstyle{\int} \bT_V, \bB\}_{BV} = V \pd_z \bB + (\pd_z V) \bB\\
	V \cdot \bPhi_r & = \{\textstyle{\int} \bT_V, \bPhi_r\}_{BV} = V \pd_z \bPhi_r + \tfrac{r}{2} (\pd_z V) \bPhi_r\\
	V \cdot \bPsi_r & = \{\textstyle{\int} \bT_V, \bPsi_r\}_{BV} = V \pd_z \bPsi_r + (1-\tfrac{r}{2})(\pd_z V) \bPsi_r\\
\end{aligned}
\ee

We see that the holomorphic translation invariance of the underlying theory becomes enhanced upon taking the $HT$ twist. 
\begin{proposition}
The $HT$ twist of a three-dimensional $\cN=2$ Chern-Simons--matter theory on $\C \times \R$ is equipped with a symmetry by the Lie algebra of holomorphic vector fields $\Vect^{hol}(\C)$.
\end{proposition}

It should be emphasized that this statement holds at the level of quantization.
Indeed, in \cite{GRWthf} it was shown that there are no one-loop anomalies to quantizing symmetries in the THF background $\C \times \R$.
This, combined with the fact that the $HT$ twist of any three-dimensional $\cN=2$ theory is one-loop exact gives the statement above at the quantum level.

\begin{proof}
It suffices to show
\begin{itemize}
\item[(1)] $\{\textstyle{\int} \bT_V, S\}_{BV} = 0$ for all holomorphic vector fields $V$, and
\item[(2)] $\{\textstyle{\int} \bT_V, \textstyle{\int} \bT_{V'}\}_{BV} = \textstyle{\int} \bT_{-[V,V']}$ for all holomorphic vector fields $V,V'$.
\end{itemize}

First, note that $\bT_V = V \bT$.
Next, since $Q \bT = \diff' \bT$ (and using that $V = V(z)\pd_z$ is holomorphic) we have
\be
	Q \bT_V = \diff' \bT_V
\ee
Assertion (1) follows by noting $Q = \{-, S\}_{BV}$.

The proof of (2) is by direct computation.
We have
\begin{align*}
	V \cdot \textstyle{\int}\bT_{V'} & = -\textstyle{\int} V' \big((V \cdot \bB) \pd_z \bA + \bB \pd_z (V \cdot \bA)\big)\\
	& \qquad + \sum\limits_r (1-\tfrac{r}{2}) \textstyle{\int} V'\big((V \cdot \bPsi) \pd_z \bPhi + \bPsi \pd_z (V \cdot \bPhi)\big)\\
	& \qquad - \sum\limits_r \tfrac{r}{2} \textstyle{\int} V'\big((V \cdot \bPhi) \pd_z \bPsi + \bPhi \pd_z (V \cdot \bPsi)\big)\\
	& = \textstyle{\int} 2 V' (\pd_z V) \bT + V' V \pd_z \bT = \textstyle{\int} \bT_{(V' \pd_z V - V \pd_z V')\pd_z}
\end{align*}
where we have used integration by parts in the last equality.
\end{proof}

We conjecture that the $HT$ twist of any three-dimensional $\cN=2$ theory equipped with $R$-symmetry has such an enhanced symmetry by holomorphic vector fields.
For those theories which are superconformal we have some strong evidence for this.
Indeed, the superconformal algebra for $\cN=2$ supersymmetry is the orthosymplectic group $\lie{osp}(2|4)$.
The $HT$ supercharge $Q_{HT}$ is a particular odd element of this Lie superalgebra and its commutant is exactly $\lie{sl}(2)$.
Enhancement for superconformal algebra asserts that the holomorphic vector fields $\del_z, z \del_z, z^2 \del_z$ generating this $\lie{sl}(2)$ can be prolongated to an action of all holomorphic vector fields.

There are other symmetries also enjoy such an enhancement. 
Consider a flavor symmetry of the chiral multiplets by a Lie group $H$ with complexified Lie algebra $\lie{h}$. 
We assume the flavor symmetry acts linearly (although this can likely be relaxed without issue) with representation matrices $(\upsilon_i)^n{}_m$; these matrices commute with the action of gauge symmetry, $[\tau_a, \upsilon_i] = 0$; and this action preserves the superpotential $(\upsilon_i \bPhi)^n \pd_{\bPhi^n} \bW = 0$. With this, it is straightforward to check that the conserved currents realizing the action of $U = U^i u_i \in \mathfrak{h} = \Lie(H)$ (the $u_i$ are a basis of $\mathfrak{f}$) are given by $\bJ_U = U^i(\bPsi \upsilon_i \bPhi)$. Just as above, we can allow the coefficients $U^i = U^i(z)$ to depend on the holomorphic coordinate $z$ without issue and find an action of $Q$ given by
\be
	Q \bJ_U = \diff' \bJ_U
\ee
and an action of holomorphic vector fields given by
\be
	V \pd_z \cdot \textstyle{\int}\bJ_U = \textstyle{\int}\bJ_{V \pd_z U}
\ee
We conclude that the twisted theory admits a natural action of holomorphic flavor transformations $\mathfrak{f} \otimes \cO_\C$, acting on the matter fields as 
\be
\begin{aligned}
	U \cdot \bPhi & = \{\textstyle{\int} \bJ_U, \bPhi\}_{BV} = U^i (\upsilon_i \bPhi)\\
	U \cdot \bPsi & = \{\textstyle{\int} \bJ_U, \bPsi\}_{BV} = -U^i (\bPsi \upsilon_i)
\end{aligned}
\ee
The following can be proved in an analogous way as in the case of holomorphic vector fields.
\begin{proposition}
Suppose that a three-dimensional $\cN=2$ theory of gauged chiral multiplets admits a flavor symmetry by a group~$H$.
Then the $Q_{HT}$-twist admits an infinitesimal symmetry by the infinite-dimensional Lie algebra~$\lie{h}\otimes \cO^{hol}(\C)$.
\end{proposition}

As we will see in the following sections, this holomorphic symmetry enhancement is not restricted to bosonic symmetries which remain after twisting.
The $HT$ twist of highly supersymmetric theories will often admit residual fermionic symmetries coming from supercharges that commute with the $HT$ supercharge $Q_{HT}$; these too admit a holomorphic enhancement.

\subsection{Descent and vacuum modules}
\label{sec:descent}

As a consequence of the symmetry enhancement results of the previous section, we obtain that the space of local operators in the $HT$ twist is naturally a representation for these enhanced algebras.

In this section we use holomorphic-topological descent to further enlarge such symmetry algebras to current algebras, where the currents are supported on two-spheres 
\beqn
S^2 \subset \C \times \R - \{0\} .
\eeqn
This situation is reminiscent of a familiar one in two-dimensional conformal field theory where the state-operator correspondence endows local operators with an action of the algebra of $S^1$-modes of the theory.
In \cite{GWrav} the first and third authors developed the resulting algebraic structure and have deemed them \textit{raviolo vertex algebras} due to their parallels with the ordinary theory of vertex algebras.
Another approach used in \cite{OhYagi,Zeng,CostelloDimofteGaiotto-boundary}, equivalent to the theory of raviolo vertex algebras, is based on $1$-shifted Poisson vertex algebras where the descent bracket is a shifted version of a $\lambda$-bracket.
We will not use the full details of raviolo vertex algebras in this note, but we briefly expound on the structure from the point of view of descent.

If the three-dimensional theory were fully topological then by Witten's topological descent the spherical modes would be labeled by the de Rham cohomology of the two-sphere $H^\bu (S^2) \simeq \C \oplus \C[-2]$.
As an example, consider the following free topological theory which consists of fields
\beqn
	\CX \in \Omega^\bu(\R^3) \otimes V, \quad \CY \in \Omega^\bu(\R^3) \otimes V^* [2]
\eeqn
where the free action is $\int \CY \d \CX$.
This is simply the Rozansky--Witten AKSZ model with target $T^*[2]V$; it is a twist of the three-dimensional $\cN=4$ hypermultiplet which we will further consider in the next section.
The local operators consist of polynomial expressions in the constant, 0-form modes of the fields $\CX,\CY$, denoted $X,Y$.
As an algebra this is
\beqn
\cO(T^* [2] V) = \Sym(V_X^* \oplus V_Y[-2]) .
\eeqn
Additionally, topological descent equips this algebra with a degree $-2$ Poisson bracket $\{-,-\}_{top}$ determined by
\beqn
\{X, Y\}_{top} = 1 .
\eeqn
There is a flavor symmetry by the Lie algebra $\lie{gl}(V)$ of matrices acting on $V$.
This algebra is a representation for $\lie{gl}(V)$ in the obvious way.
Via topological descent, this symmetry is enhanced to a symmetry by the graded Lie algebra
\beqn
\lie{gl}(V) \otimes H^\bu (S^2) = \lie{gl}(V) \ltimes \lie{gl}(V)[-2] .
\eeqn
The currents for the original $\lie{gl}(V)$ symmetry are given by 
\beqn
-\oint_{S^2} \CY A \CX , \quad A \in \lie{gl}(V) .
\eeqn
Via the shifted descent bracket, this current acts on a local operator $O$ through the quadratic local operator $M_A = -Y (A X)$:
\beqn
A \in \lie{gl}(V) \colon O \mapsto \{M_A, O\}_{top} .
\eeqn
The currents for the additional copy of $\lie{gl}(V)[-2]$ sitting in cohomological degree $+2$ are
\beqn
-\oint_{S^2} \CY A \CX \; \text{dvol}_{S^2} , \quad A \in \lie{gl}(V)[-2] ,
\eeqn
where $\text{dvol}_{S^2}$ stands for the volume element of the two-sphere.
Notice that such currents only depend on the lowest form components of $Y,X$, which reveals that its cohomological degree is $+2$ as expected.
At the level of local operators this acts simply by inserting the quadratic local operator $M_B$:
\beqn
B \in \lie{gl}(V)[-2] \colon \cO \mapsto M_B \cO .
\eeqn
Notice that if $B,B' \in \lie{gl}(V)[-2]$ then the relation $[B,B'] = 0$ holds since the algebra of local operators is commutative in cohomology.

In the situation of the $HT$ twist, the currents must be sensitive to the THF structure on $\C \times \R$.
To obtain algebraic models for the enhanced symmetry algebras we will consider punctured affine space instead of a two-sphere of a fixed radius; of course these spaces are homotopy equivalent, so in the topological situation it does not change anything.
Being a submanifold of $\C \times \R$, once-punctured affine space is naturally equipped with a THF structure.
In particular we can consider the commutative dg algebras
\beqn
\cA^{(j),\bu} \define \bOmega^{(j),\bu} (\C \times \R - \{0\})
\eeqn
where, in practice, $j \in \frac12 \Z$.
This complex is equipped with the differential $\d'$ which is the restriction of the standard one on $\C \times \R$ defined by $\d' = \d \zbar \del_{\zbar} + \d t \del_t$.
When $j = 0$ the degree zero part of this complex simply consists of smooth functions on $\C \times \R - \{0\} = \R^3 - \{0\}$ and the zeroth cohomology is the algebra of smooth functions which are flat along the foliation determined by the THF structure.

As a remark, we point out that the ordinary de Rham complex of $\C \times \R - \{0\}$ can be written in terms of the complexes above as
\beqn
\bOmega^\bu(\C \times \R - \{0\}) = \cA^{(0),\bu} \xto{\del} \cA^{(1),\bu}[-1] 
\eeqn
where $\del$ is locally given by $\del = \d z \del_z$.
From this we observe parallels with the usual Hodge decomposition of the de Rham cohomology of $\P^1$ using its complex structure.
The complexes we consider here are distinct from the usual Dolbeault complexes of $\P^1$ as we utilize the inherented THF structure on $S^2$ as a submanifold of $\C \times \R$ rather than its complex structure.

Any $S^2$ current can be written as an observable on the fields of a three-dimensional $\cN=2$ theory evaluated on $\C \times \R - \{0\}$.
The complex $\cA^{(0),\bu}$ is equipped with a version of the residue pairing which is simply given by integration along any two-sphere
\beqn
\oint_{S^2} \alpha(z,t) \wedge \d z , \quad \alpha \in \cA^{(0),1}
\eeqn
Note that this expression is only nonzero when $\alpha$ is a one-form in the THF complex.
There is a particular one-form that we denote by
\beqn\label{eqn:omega}
\omega = c \frac{2 \zbar \d t - t \d \zbar}{r^3} \in \cA^{(0),1}
\eeqn
where the constant $c$ is normalized so that $\oint_{S^2} \omega \wedge \d z = 1$.
The element $\omega$ is easily seen to be $\d'$-closed and hence represents a cohomology class (see below).
The two-form $\omega \wedge \d z$ is an integral kernel for the distributional operator $(\d')^{-1}$; therefore it plays the role of the propagator for an $HT$ twisted theory.
In \cite{GRWthf} the third author used a regularized version of this propagator to study renormalization for $HT$ twisted theories.

The $\d'$-cohomology $H^{(j),\bu}(\C \times \R - \{0\})$ of the complex $\cA^{(j),\bu}$ is explicit to describe.
When $r=0$ the cohomology is concentrated in degrees zero and one.
In degree zero there is an isomorphism
\beqn
H^{(0),0} (\C \times \R - \{0\}) \simeq \cO^{hol}(\C) 
\eeqn
which is induced by pulling back holomorphic functions along $\C \times \R - \{0\} \to \C$.
In degree one there is a dense embedding 
\beqn
\C[\del_z] \omega \hookrightarrow H^{(0),1} (\C \times \R - \{0\})
\eeqn
where $\omega$ is as in \eqref{eqn:omega}.
The vector space $\C[\del_z]$ is a module over $\cO^{hol}(\C)$ by the rule that $z^n \cdot \del_z^m \omega = \frac{m!}{(m-n)!} \del_z^{m-n} \omega$ if $m \geq n$ and zero otherwise.

Now, consider a supersymmetric flavor symmetry by some group $H$ on a three-dimensional $\cN=2$ theory.
In the previous section we have seen how the infinitesimal symmetry is enhanced from the Lie algebra $\lie{h} = \op{Lie}(H)$ to the current algebra $\lie{h} \otimes \cO^{hol}(\C)$.
Holomorphic-topological descent further enhances this symmetry to the dg Lie algebra
\beqn
\lie{h} \otimes \cA^{(0),\bu} .
\eeqn
The differential is $\id_{\lie{h}} \otimes \d'$ and the bracket is determined by the bracket on $\lie{h}$ together with the product on $\cA^{(0),\bu}$.
This is the three-dimensional $\cN=2$ $HT$ twisted analog of the mode algebra $\lie{h} [z,z^{-1}]$ in a chiral CFT.
In cohomology we obtain a symmetry by the graded Lie algebra
\beqn\label{eqn:fcoh}
\lie{h} \otimes H^{(0),\bu} (\C \times \R - \{0\}) \simeq \lie{h} \otimes \cO^{hol}(\C) \ltimes \lie{h} \otimes \C[\del_z] \omega [-1] .
\eeqn
The semi-direct product utilizes the $\cO^{hol}(\C)$-module structure on $\C[\del_z]$ defined above.

Denote the space of local operators in the $HT$ twist of a three-dimensional $\cN=2$ theory by $\op{Ops}$.
Such local operators form a commutative algebra and the cohomology is equipped with a shifted $\lambda$-bracket of cohomological degree $-1$ that we denote by $\htl -, -\htr^{(n)}$, $n \geq 0$ \cite{OhYagi,CostelloDimofteGaiotto-boundary}.
Explicitly, the bracket between two local operators is defined by
\beqn
\htl \cO, \cO'\htr^{(n)} (w,s) \define \oint_{S^2} z^n \d z \; \til \cO^{(1)} (z,t) \cO'(w,t)
\eeqn
where $\cO^{(1)}(z,t) \in \cA^{(0),1} \otimes \Ops$ is the one-form holomorphic-topological descendant of the local operator $\cO(z,t)$.

For a concrete example, we consider the the $HT$ twist of the theory of a free chiral multiplet with values in a vector space $V$ (we will not need to be specific about $R$-symmetry in what follows).
The cohomology of the algebra of local operators $\Ops$ is freely generated by even symbols $\del^n_z \phi$ of cohomological degree zero and odd symbols $\del^m_z \psi$ of cohomological degree $+1$ where $n,m \geq 0$.

As in the topological example above, we contemplate the flavor symmetry by the Lie algebra $\lie{gl}(V)$ and its enhancement to
\beqn\label{eqn:glVht}
\lie{gl}(V) \otimes H^{(0),\bu}(\C \times \R - \{0\}) = \lie{gl}(V) \otimes \cO^{hol}(\C) \ltimes \lie{gl}(V) \otimes \C[\del_z] \omega [-1] .
\eeqn
We have already pointed out how the degree zero cohomology classes $A \otimes z^n \in \lie{gl}(V) \otimes \cO^{hol}(\C)$ give rise to currents
\beqn
\oint_{S^2} z^n \d z \, \psi A \phi .
\eeqn
At the level of local operators, such elements act through the HT $\lambda$-bracket
\beqn
A \otimes z^n \colon O \mapsto \htl \psi A \phi, O \htr^{(n)} .
\eeqn

Next, consider the classes in \eqref{eqn:glVht} of degree $+1$ which all have the form $B \otimes \del_z^n \omega$.
In the twist of the free chiral multiplet, the currents associated to such classes have the form
\beqn
\oint_{S^2} \del_z^n \omega \d z \, \psi A \phi .
\eeqn
At the level of local operators, such elements act via `creation' operators as
\beqn
B \otimes \del_z^n \omega \colon O \mapsto \del_z^n(\psi B \phi) O .
\eeqn

A totally analogous symmetry enhancement holds for the (twisted) superconformal algebra.
Above, we have seen that the $HT$ twist of a three-dimensional superconformal theory has enhanced symmetry by the Lie algebra $\text{Vect}^{hol}(\C)$.
This further enhances to a symmetry by the following dg Lie algebra.
Let $T^{hol}$ denote the complex rank one bundle on $\C \times \R$ locally spanned by the vector field $\del_z$.
The sheaf of sections of $T^{hol}$ which are holomorphic in the $z$-direction (closed for the $\d'$-operator) is endowed with a Lie bracket, and globally on $\C \times \R$ this recovers the Lie algebra of holomorphic vector fields.
Further, the complex of sheaves $\Omega^{(-1),\bu}$ on $\C \times \R$ is a free resolution of this sheaf and can be given the structure of a sheaf of dg Lie algebras.
The dg Lie algebra
\beqn
\cA^{(-1),\bu} \define \Omega^{(-1),\bu} (\C \times \R - \{0\})
\eeqn
is the desired enhancement of the Lie algebra $\Vect^{hol} (\C)$.
Indeed, the zeroth cohomology of this complex is exactly the Lie algebra of holomorphic vector fields.
The first cohomology is nontrivial and generated by classes of the form $(\del_z^n \omega) \del_z$ where $n \geq 0$.
The higher cohomology vanish.
The dg Lie algebra $\cA^{(-1),\bu}$ is the twisted three-dimensional analog of the Virasoro algebra (really the Witt algebra, as we have not included a central extension).

We conclude this subsection with a remark that this discussion applies at the level of cohomology.
In the case of a flavor symmetry labeled by $\lie{h}$, we have seen how to construct in $Q_{HT}$-cohomology, a symmetry by the graded Lie algebra \eqref{eqn:fcoh}.
This cohomology loses information that is present at the cochain level. 
Indeed, there are higher $L_\infty$ operations present in cohomology reflecting the fact that the dg Lie algebra $\lie{h} \otimes \cA^{(0),\bu}$ is not formal.
Consequently, the cohomology of local operators in the $HT$-twist will be enriched to $L_\infty$ modules for this $L_\infty$ algebra.
We do not characterize this structure here.

\subsection{The perspective of factorization algebras}

Finally, we remark on another approach to the theory of observables for the $HT$ twist of a three-dimensional supersymmetric theory based on the theory of factorization algebras.
A central result in \cite{CG,CG2} is that the observables of any quantum field theory have the structure of a factorization algebra.
Here it is necessary that observables can be evaluated on an arbitrary open set of the spacetime manifold $U \mapsto \op{Obs}(U)$.
For two open sets $U,V$ which are disjoint and lie within a bigger open set $W$ part of the structure of a factorization algebra is a sort of multiplication map
\beqn
\op{Obs}(U) \otimes \op{Obs}(V) \to \op{Obs}(W) 
\eeqn 
which satisfies a sort of associativity axiom.
We understand local operators supported at a point (denoted $\op{Ops}$ above) as observables supported on an arbitrarily small open set containing the point.
There are higher-ary operations, generalizing the multiplication above to arbitrary numbers of inputs of disjoint open sets, resulting in a structure similar to that of an operad.
We refer to \cite{CG} for more precise definitions.\footnote{We have only mentioned the structure of a \textit{pre}factorization algebra. 
A factorization algebra is one which also satisfies a gluing axiom which we will not utilize in this paper.}

In a three-dimensional \textit{topological} theory the factorization algebra of observables result in a familiar operadic structure called an $E_3$ algebra---an algebra over the operad of little three-disks.
In cohomology this recovers the ($2$-shifted) topological descent bracket.
Similarly, the factorization algebra encoding a three-dimensional holomorphic-topological theory results in the $1$-shifted $\lambda$-bracket that we just recollected.
Sometimes it is convenient to work at the full level of the $HT$-twisted factorization algebra, which we comment on below.

\section{$\mathcal{N}=4$ Yang-Mills Gauge Theories}
\label{sec:SYM}

In this section we consider the simple example of $\CN=4$ hypermultiplets gauged with $\CN=4$ vector multiplets. We assume that the hypermultiplets transform in the complex symplectic representation $\CR$ of the (complexified) gauge group $G$.

The $HT$-twisted theory consists of an $\CN=2$ vector multiplet $(\bA, \bB)$, an $\CN=2$ adjoint-valued chiral multiplet $(\bPhi, \bLambda)$ of $R$-charge $1$, and an $\CR$-valued chiral multiplet $(\bZ, \bPsi)$ of $R$-charge $\tfrac{1}{2}$. In addition, the theory has a superpotential of the form $\bW = -\tfrac{1}{2} \bPhi^a (\tau_a)_{mn} \bZ^m \bZ^n$, where $(\tau_a)_{mn} = \Omega_{ml} (\tau_a)^{l}{}_n = (\tau_a)_{nm}$, i.e. the superpotential is (minus) the pairing of $\bPhi$ and the moment map $\bnu$ for the $G$ action on $\CR$. The $HT$-twisted action is then given by
\be
\label{eq:twistedactionSYM}
S = \int \bB F'(\bA) + \bLambda \diff'_{\bA} \bPhi + \bPsi \diff'_{\bA} \bZ - \bnu \bPhi,
\ee
and the action of $Q_{HT}$ is given by 
\be
\begin{aligned}
	\label{eq:QSYM}
	Q \bA & = F'(\bA) \qquad & Q \bB & = \diff'_{\bA} \bB - \bmu\\
	Q \bPhi & = \diff'_{\bA} \bPhi & Q \bLambda & = \diff'_{\bA} \bLambda - \bnu\\
	Q \bZ & = \diff'_{\bA} \bZ \qquad & Q \bPsi & = \diff'_{\bA} \bPsi - \bPhi \bZ\\
\end{aligned}
\ee
We note that the (modified, extended) stress tensor is given by
\be
\bT_V = \iota_V \bigg(- \bB \pd \bA + \tfrac{1}{2}\big(\bLambda \pd \bPhi - \bPhi \pd \bLambda\big) + \tfrac{1}{4}\big(3 \bPsi \pd \bZ - \bZ \pd \bPsi\big)\bigg) .
\ee

\subsection{Supersymmetric extension of $\textrm{Vect}(\C)$}

The $\cN=4$ superconformal algebra is $\lie{osp}(4|4)$ and the $HT$ supercharge $Q_{HT}$ lives inside the $\cN=2$ subalgebra $\lie{osp}(2|4) \subset \lie{osp}(4|4)$.
The $HT$ twist of the $\cN=4$ superconformal algebra is $\lie{osp}(2|2)$.
The even part of this algebra is $\lie{gl}(1) \times \lie{sl}(2)$.
As in the $\cN=2$ case, the $\lie{sl}(2)$ acts geometrically through the vector fields $\del_z,z \del_z,z^2 \del_z$.
The $\lie{gl}(1)$ is the stabilizer of $Q_{HT}$ in the $\cN=4$ $R$-symmetry algebra~$\lie{so}(4)$.
We will see explicitly how this symmetry gets enhanced to an infinite-dimensional symmetry algebra $\lie{a}_{\cN=4}$ which turns out to be the positive part of the $\cN=2$ super Virasoro algebra $\lie{vir}_{\cN=2}$ of type Neveu--Schwarz.

It is straightforward to see that every model described above admits a $\C^\times$ flavor symmetry under which $\bZ$ transforms with weight $1$ and $\bPhi$ transforms with weight $-2$; the current generating the (holomorphically extended) symmetry is simply
\be
\bJ_S = S \big(\bPsi \bZ - 2 \bLambda \bPhi\big)
\ee
where $S = S(z)$ need not be constant. The constant part of the bosonic symmetry generated by this current is the remnant of the $\CN=4$ $R$-symmetry group that preserves~$Q_{HT}$.

In addition, there are always two local operators
\be
	\bTheta_{\Gamma} = \Gamma \big(\tfrac{1}{2} \Omega^{-1}(\bPsi, \bPsi) - \bB \bPhi\big) \qquad \wt{\bTheta}_{\wt{\Gamma}} = \wt{\Gamma}\big(\tfrac{1}{2} \Omega(\bZ, \pd \bZ) + \bLambda \pd \bA\big)
\ee
that generate fermionic symmetries parameterized by $\Gamma = \Gamma(z)$ and $\wt{\Gamma} = \wt{\Gamma}(z)$ of the $HT$ twisted action. On the twisted superfields, we find that the action of $\Gamma$ takes the form
\be
\begin{aligned}
	\Gamma \cdot \bA & = \Gamma \bPhi \qquad & \Gamma \cdot\bB & =0\\
	\Gamma \cdot \bPhi & = 0 \qquad & \Gamma \cdot\bLambda & = \Gamma \bB\\
	\Gamma \cdot \bZ & = \Gamma \Omega^{-1}(\bPsi, -) \qquad & \Gamma \cdot\bPsi & = 0\\
\end{aligned}
\ee
and the action of $\wt{\Gamma}$ is given by
\be
\begin{aligned}
	\wt{\Gamma} \cdot \bA & = 0 \qquad & \wt{\Gamma} \cdot \bB & = -\wt{\Gamma} \pd \bLambda - (\pd\wt{\Gamma})\bLambda\\
	\wt{\Gamma} \cdot \bPhi & = -\wt{\Gamma} \pd \bA \qquad & \wt{\Gamma} \cdot \bLambda & = 0\\
	\wt{\Gamma} \cdot \bZ & = 0 \qquad & \wt{\Gamma} \cdot \bPsi & = \wt{\Gamma} \Omega(\pd \bZ, -) + \tfrac{1}{2} (\pd\wt{\Gamma}) \Omega(\bZ,-)\\
\end{aligned}
\ee

The action of $V\pd_z$ on the integrals $\int\bTheta_{\Gamma}$ and $\int\wt{\bTheta}_{\wt{\Gamma}}$ encodes the commutators of $V\pd_z$ and $\Gamma, \wt{\Gamma}$:
\be
	V\pd_z \cdot \textstyle{\int} \bTheta_{\Gamma} = \textstyle{\int}\bTheta_{-V \pd_z \Gamma + \scriptsize{\frac{1}{2}} (\pd_z V)\Gamma} \qquad \qquad V\pd_z \cdot \textstyle{\int}\wt{\bTheta}_{\wt{\Gamma}} = \textstyle{\int}\wt{\bTheta}_{-V \pd_z\wt{\Gamma} + \scriptsize{\frac{1}{2}} (\pd_z V)\wt{\Gamma}}
\ee
so that
\be
	[V\pd_z, \Gamma] = V \pd_z \Gamma - \tfrac{1}{2} (\pd_z V) \Gamma \qquad [V\pd_z, \wt{\Gamma}] = V \pd_z \wt{\Gamma} - \tfrac{1}{2} (\pd_z V) \wt{\Gamma}
\ee
so that $\Gamma$ and $\wt{\Gamma}$ transform as sections of $K^{-1/2}_\C$. In a similar fashion, the action of $S$ encodes the commutators of $S$ and $\Gamma, \wt{\Gamma}$:
\be
	S \cdot \textstyle{\int}\bTheta_{\Gamma} = \textstyle{\int}\bTheta_{-2 S\Gamma} \qquad \qquad S \cdot \textstyle{\int}\wt{\bTheta}_{\wt{\Gamma}} = \textstyle{\int}\wt{\bTheta}_{2 S \wt{\Gamma}}
\ee
which implies
\be
	[S, \Gamma] = 2 S\Gamma \qquad [S, \wt{\Gamma}] = - 2 S\wt{\Gamma}
\ee
Finally, we find that action of $\Gamma$ on $\int \wt{\bTheta}_{\wt{\Gamma}}$ encodes the (anti)commutator of $\Gamma$ and $\wt{\Gamma}$
\be
	\Gamma \cdot \textstyle{\int}\wt{\bTheta}_{\wt{\Gamma}} = \textstyle{\int}\bT_{-\Gamma \wt{\Gamma}} + \textstyle{\int}\bJ_{\scriptsize{\frac{1}{4}}(\wt{\Gamma} \pd_z \Gamma - \Gamma \pd_z \wt{\Gamma})}
\ee
corresponding to
\be
	[\Gamma, \wt{\Gamma}] = \overset{K^{-1}_\C = \op{Vect}(\C)}{\overbrace{(\Gamma \wt{\Gamma})}} + \tfrac{1}{4}\overset{\CO_\C}{\overbrace{(\Gamma \pd_z \wt{\Gamma} - \wt{\Gamma}\pd_z\Gamma)}}
\ee
We will denote by $\lie{a}_{\CN=4}$ the algebra generated by the bosonic generators $V\pd_z, S$ together with the fermionic generators $\Gamma, \wt{\Gamma}$. Putting this together, we obtain the following result.

\begin{proposition}
The $HT$ twist of a three-dimensional $\CN=4$ theory built from coupling an arbitrary number of vector multiplets and hypermultiplets admits a natural action of a Lie superalgebra $\lie{a}_{\CN=4}$.
\end{proposition}

If $\Gamma$ and $\wt{\Gamma}$ are constant, so that $\pd_z \Gamma = \pd_z \wt{\Gamma} = 0$, we see that these two currents bracket to the constant vector field $\pd_z$ generating holomorphic translations. Indeed, the fermionic symmetries generated by these currents correspond to (the holomorphic enhancement of) the supercharges that deform the $HT$ twist to the topological $A$ and $B$ twists -- deforming the $HT$ twisted action by $\int \bTheta_1$ (resp. $\int \wt{\bTheta}_1$) results in the $A$ twist (resp. $B$ twist).%
\footnote{One should also replace the stress tensor $\bT_{V}$ by $\bT_V \pm \tfrac{1}{4}\big(\pd_z \bJ_V - \bJ_{\pd_z V}\big)$ in accordance with the twisted spins of the $A$ ($-$) and $B$ ($+$) twists. We also note that the cohomological grading $C$ is altered in passing to the $A$ and $B$ twists.} %
The remaining fermionic generators in the $HT$ twist of the $\CN=4$ superconformal algebra are $\Theta_z$ and $\wt{\Theta}_z$.

We expect, but do not prove here, that the $HT$ twist of a general $\cN=4$ theory (with $R$-symmetry) admits an action by $\lie{a}_{\cN=4}$. 
For example, the dimensional reduction of the holomorphic twist of a theory of class $S$ should admit such a symmetry.

Also, note that $\lie{a}_{\cN=4}$ is the positive part (in the sense of Fourier modes) of the chiral sector of the $\cN=2$ super Virasoro algebra
\be
	\lie{a}_{\CN=4} = \lie{vir}_{\CN=2}^{\geq 0}\,.
\ee

The action of $\lie{a}_{\CN=4}$ on an the HT twist of an arbitrary $\cN=4$ theory implies there are various deformations of the $HT$-twist given by square-zero fermionic symmetries. These deformations arise adding the corresponding term to the action, e.g. $S \to S + \int\! \Theta_{\Gamma}$. We now outline some simple cases:
\begin{itemize}
	\item The element $\Gamma = 1$ gives the $A$-twist of the three-dimensional $\cN=4$ theory as a further deformation of the HT twist and corresponds to the supercharge $Q_-^{-\dot{+}}$ of the supersymmetry algebra.
	The $A$-twisted action takes the form
	\be
	\begin{aligned}
		S_A & = S + \textstyle{\int} \bTheta_1\\
		& = \int \bB F'(\bA) + \bLambda \diff'_{\bA} \bPhi + \bPsi \diff'_{\bA} \bZ - \bnu \bPhi + \tfrac{1}{2} \Omega^{-1}(\bPsi, \bPsi) - \bB \bPhi
	\end{aligned}
	\ee
	cf. Section 4.1 of \cite{twistedN=4}.
	The cohomology of $\lie{a}_{\cN=4}$ with respect to this element is trivial.
	\item The $\Gamma = z$ is a superconformal deformation and is equivalent placing the $B$-twist of the original supersymmetric theory in the $\Omega$ background.
	The cohomology of $\lie{a}_{\cN=4}$ with respect to this element is also trivial.
	\item More generally, we can consider deforming by the element $\Gamma = z^l$ for $l \geq 2$.
	The cohomology of $\lie{a}_{\cN=4}$ with respect to this nilpotent element has $l-1$ bosonic generators and $l-1$ fermionic generators; the bosonic subalgebra is identified with holomorphic vector fields vanishing at $z=0$ modulo the ideal generated by $z^l \pd_z$. 
\end{itemize}
Similarly, trading $\Gamma$ for $\til \Gamma$ in the above two items corresponds to flipping $A$ and $B$.

The algebra $\lie{a}_{\cN=4}$ is equipped with a $\mathbb{Z}_2$ outer automorphism defined by
\begin{align*}
	S & \leftrightarrow -S \\
	\Gamma & \leftrightarrow \til \Gamma .
\end{align*}
This lifts the mirror automorphism of the three-dimensional $\CN=4$ supersymmetry algebra that exchanges the Higgs and Coloumb branch $R$-symmetries. We expect that the $HT$ twists of mirror $\CN=4$ theories will be identified in a way that intertwines their actions of $\lie{a}_{\CN=4}$ with this automorphism. For example, the deforming by $\Theta_{z^l}$ on one side of mirror symmetry should be equivalent to deforming by $\wt{\Theta}_{z^l}$ in the other side. This generalizes the usual statements about exchanging the $A$ and $B$ twists (for $l = 0$) and their Omega-backgrounds (for $l = 1$).

\subsection{$\CN=4$ flavor symmetries}
\label{sec:N=4flavor}
A natural question to ask is what the above analysis implies for $\mathcal{N}=4$ flavor symmetries. Although we will focus on Higgs-branch flavor symmetries, there are analogous consequences for Coulomb branch flavor symmetries.

Suppose a second (complex reductive) group $H$ acts (linearly) on $\CR$ in a way that preserves the symplectic form $\Omega$ and commutes with the action of the gauge group $G$; let $(\upsilon_i)^{n}{}_m$ be representation matrices for the action of $H$. It immediately follows that the currents $\bL_U = U^i (\bPsi \upsilon_i \bZ)$ generate a flavor symmetry of the $HT$-twisted theory. Moreover, they are chargeless (weight $0$) under the remnant $R$-symmetry $S$ and annihilated by the odd generator $\Gamma$:
\be
	S \cdot \bL_U = 0 \qquad  \Gamma \cdot \bL_U = 0
\ee
On the other hand, the action of $\wt{\Gamma}$ is related to the flavor symmetry moment map $\bXi_{\Upsilon} = \tfrac{1}{2} \Upsilon^{i} (\upsilon_i)_{nm}\bZ^n \bZ^m$ via
\be
	\wt{\Gamma} \cdot \textstyle{\int} \bL_U = \textstyle{\int}\bXi_{-\wt{\Gamma} \pd U}
\ee

The bosonic operator $\bXi_\Upsilon$ has weight $2$ under the remnant $R$-symmetry and brackets trivially with $\wt{\bTheta}_{\wt{\Gamma}}$:
\be
	S \cdot \textstyle{\int}\bXi_\Upsilon = \textstyle{\int}\bXi_{-2 S \Upsilon} \qquad \wt{\Gamma}\cdot \textstyle{\int}\bXi_\Upsilon = 0
\ee
The action of $\Gamma$ takes the following simple form:
\be
	\Gamma \cdot \textstyle{\int}\bXi_{\Upsilon} = \textstyle{\int}\bL_{\Gamma \Upsilon}
\ee
We also note that the action $V \pd_z$ is given by
\be
	V \pd_z \cdot \textstyle{\int}\bXi_{\Upsilon} = \textstyle{\int}\bXi_{-V\pd_z\Upsilon - \scriptsize{\frac{1}{2}}(\pd_z V) \Upsilon}
\ee
and the action of $U$ is given by
\be
	U \cdot \textstyle{\int}\bXi_{\Upsilon} = \textstyle{\int}\bXi_{-[U,\Upsilon]}
\ee

All together, we see that the existence of an $H$ Higgs branch flavor symmetry implies the following.

\begin{proposition}
Suppose a three-dimensional $\cN=4$ theory of vector multiplets and hypermultiplets admits a Higgs-branch flavor symmetry by a group $H$.
The $HT$ twisted theory admits an action of a Lie superalgebra $\lie{f}_{H,\CN=4}$ defined by
\be
\lie{f}_{H,\CN=4} = \overset{U}{\overbrace{\CO_\C \otimes \mathfrak{h}_0}} \oplus \Pi \overset{\Upsilon}{\overbrace{K^{1/2}_\C \otimes \mathfrak{h}_2}}
\ee
where the subscripts denote the transformation properties with respect to the bosonic subalgebra $(\lie{a}_{\CN=4})^+$. 
The Lie bracket on the bosonic subalgebra $(\lie{f}_{H,\CN=4})^+$ corresponds to the natural Lie bracket on $\CO_\C \otimes \mathfrak{h}$; the action of the bosonic subalgebra $(\lie{f}_{H,\CN=4})^+$ on the fermionic subspace $(\lie{f}_{H,\CN=4})^-$ uses the adjoint action of $\mathfrak{h}$ on itself and the action of functions $\CO_\C$ on sections of $K^{1/2}_\C$; and the bracket of two fermionic elements vanishes.
\end{proposition}

As in the case of the enhanced superconformal algebra $\lie{a}_{\cN=4}$ we can understand further deformations of the enhanced flavor symmetry algebra $\lie{f}_{H, \cN=4}$ by square-zero elements in $\lie{a}_{\cN=4}^-$.
\begin{itemize}
	\item
	The cohomology of $\lie{f}_{H, \cN=4}$ with respect to the element $\Gamma=1$ is trivial.
	In fact, after applying the $A$-twisting homomorphism the dg Lie algebra $(\lie{f}_{H, \cN=4}, \Gamma = 1)$ is isomorphic to the de Rham Lie algebra $(\lie{f}_{H, \cN=4})_{dR}$, consistent with the fact that Higgs branch symmetries do not act on the Coloumb branch chiral ring.%
	\footnote{We note that, although Higgs branch symmetries do not act on the Coulomb branch, mass parameters for Higgs branch symmetries can be used to deform/resolve the Coulomb branch chiral ring. The complex mass parameters arise by deforming by a fermionic element in ($\lie{f}_{H, \cN=4})_{dR}$; they are not Maurer-Cartan elements, instead leading to an equivariant differential.} %
	\item The element $\til \Gamma = 1$ leads to the $B$-twist of the three-dimensional $\cN=4$ supersymmetric theory.
	The cohomology of $\lie{f}_{H, \cN=4}$ with respect to this element is isomorphic to the Lie algebra $\lie{h}$, the Lie algebra of the flavor group we started with.
	This is consistent with the fact that Higgs branch symmetries act on the Higgs branch.
	\item The element $\Gamma = z$ leads to a superconformal deformation and is equivalent to placing the $B$-twist of the original supersymmetric theory in an $\Omega$-background. 
	The cohomology of $\lie{f}_{H, \cN=4}$ with respect to this element is isomorphic to the Lie algebra $\lie{h}$. This should be identified with a quantization of the symmetry in the previous item.
	\item The deformation by $\til \Gamma = z$ is another superconformal deformation and is equivalent placing the $A$-twist of the original supersymmetric theory in the $\Omega$-background.
	The cohomology of $\lie{f}_{H, \cN=4}$ with respect to this element is isomorphic to the semi-direct product Lie superalgebra $\lie{h} \ltimes \Pi \lie{h}$. The odd part of this cohomology yields further deformations corresponding to turning on a complex mass deformation.
	
\end{itemize}

To witness symmetries of the Coloumb branch of the three-dimensional $\cN=4$ theory one switches the roles of $\Gamma$ and $\til \Gamma$ (and $S$ with $-S$). In the above $\CN=4$ theory, the portion of these symmetries visible at the level of the action is called the topological flavor symmetry and given by the Pontrjagin dual to $\pi_1(G)$. The supersymmetric extension of this symmetry is generated by the local operators
\be
\wt{\bL}_{\wt{U}} = \wt{U} \pd \Tr (\bA)\,, \qquad  \wt{\bXi}_{\wt{\Upsilon}} = \wt{\Upsilon} \Tr(\bPhi)\,.
\ee

\subsection{Example: superconformal deformation of an $HT$-twisted hypermultiplet}
\label{sec:hyperSC}
To make the above discussions more explicit, we consider the case of a free hypermultiplet. The HT twisted theory has two bosonic fields $\bZ^\alpha$ and two fermionic fields $\bPsi_\alpha$ with BV/BRST differential given by
\be
Q \bZ = \diff' \bZ \qquad Q \bPsi = \diff' \bPsi.
\ee
From this, we can read off the algebra of local operators in the HT twist: only the lowest form components are $Q$-closed, and their $\bar{z},t$ dependence is exact; we are left with a commutative vertex algebra with four generators: $Z^\alpha, \psi_\alpha$. There is a $1$-shifted Poisson structure on this algebra, determined from holomorphic-topological descent:
\be
\{\!\{Z^\alpha, \psi_\beta\}\!\} = \delta^\alpha{}_\beta
\ee

The local operators generating the flavor symmetry action can be read off of the above. Namely, we consider the following local operators:
\be
L^{\alpha}{}_\beta = \psi_\beta Z^\alpha \qquad \xi^{\alpha \beta} = Z^\alpha Z^\beta
\ee

We now consider the superconformal deformation by the Hamiltonian for the odd symmetry $\Gamma = z$ in the algebra $\lie{a}_{\cN=4}$, i.e. we add $\int \bTheta_{\Gamma=z}$ to the action. This deformation corresponds to placing the $B$-twist of the original supersymmetric theory in the $\Omega$-background. This modification changes the action of the BV/BRST supercharge to
\be
Q_{SC} \bZ^\alpha = \diff' \bZ^\alpha + z \Omega^{\beta \alpha} \bPsi_\beta \qquad Q_{SC} \bPsi_\alpha = \diff' \bPsi_\alpha
\ee
We can analyze the vectorspace of local operators via a spectral sequence, where the first step takes the cohomology with respect to $\diff'$, again restricting us to local operators built as polynomials in the holomorphic derivatives in $Z^\alpha, \psi_\alpha$. The differential on the second page corresponds to the action of $\Gamma=z$ and can be identified with taking the descent bracket with the Hamiltonian $\frac{z}{2}\Omega^{-1}(\psi, \psi)$:
\beqn
\{\!\{\tfrac{z}{2} \Omega^{-1}(\psi, \psi), Z^\alpha(z) \}\!\} = z \Omega^{\beta \alpha} \psi_\beta(z) 
\eeqn
and 
\beqn
\{\!\{\tfrac{z}{2} \Omega^{-1}(\psi, \psi), \psi^\alpha(z) \}\!\} = 0.
\eeqn
Thus, local operators in this superconformal deformation are supported at $z=0$, independent of~$t$, and generated by $Z^\alpha(0)$.

\begin{figure}
	\centering
	\begin{tikzpicture}
		\node (int) at (0,-1) {$\otimes$};
		\node (Z1) at (-3.5,0) {$Z^{\alpha}(0,0)$};
		\node (bZ1) at (-2.5,0) {$\bullet$};
		\node (Z2) at (-3.5,-2) {$Z^{\beta}(0,-T)$};
		\node (bZ2) at (-2.5,-2) {$\bullet$};
		\draw[middlearrow={latex}, thick] (bZ1.center) to [in=135, out=0] (int.center);
		\draw[middlearrow={latex}, thick] (bZ2.center) to [in=225, out=0] (int.center);
		\draw[dashed] (-2.5,2.5) -- (-2.5,-2.5);
	\end{tikzpicture} \hspace{1cm} \raisebox{2.5cm}{$-$} \hspace{1cm} \begin{tikzpicture}
		\node (int) at (0,1) {$\otimes$};
		\node (Z1) at (-3.5,0) {$Z^{\alpha}(0,0)$};
		\node (bZ1) at (-2.5,0) {$\bullet$};
		\node (Z2) at (-3.5,2) {$Z^{\beta}(0,T)$};
		\node (bZ2) at (-2.5,2) {$\bullet$};
		\draw[middlearrow={latex}, thick] (bZ1.center) to [in=225, out=0] (int.center);
		\draw[middlearrow={latex}, thick] (bZ2.center) to [in=135, out=0] (int.center);
		\draw[dashed] (-2.5,2.5) -- (-2.5,-2.5);
	\end{tikzpicture}
	\caption{The difference of Feynman diagrams computing the commutator $[Z^\alpha, Z^\beta]$. The dotted line is the $z = 0$ axis and the interaction vertex is $\int \frac{z}{2} \Omega^{-1}(\bPsi, \bPsi)$.}
	\label{fig:hyperquantization}
\end{figure}
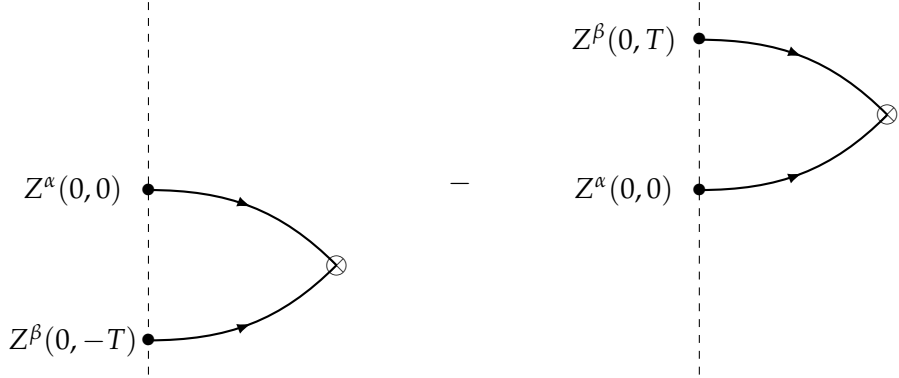

Not only does this deformation reduce us to local operators built from the $Z^\alpha$ placed at $z = 0$, it introduces a non-commutativity controlled by the symplectic form $\Omega$. It is customary to dress the interaction vertex by a ``quantization parameter'' $8\pi \varepsilon \in \C$, with a convenient choice of normalization factor.%
\footnote{We note that the $\Omega$-background quantization parameter $\varepsilon$ should not be conflated with Planck's constant $\hbar$, so we choose to denote it by a different character.} %
In order to extract this non-commutativity, we consider two configurations of insertions: the first (resp. second) configuration has an insertion of $Z^\alpha$ at $z = 0$ and $t = 0$ and an insertion of $Z^\beta$ at $z = 0$ and $t = T$ (resp. $t = - T$), for $T > 0$. The difference of the corresponding two-point functions will be our measure of the resulting non-commutativity. This difference of two-point functions is given by the difference the two Feynman amplitudes illustrated in Fig. \ref{fig:hyperquantization}; explicitly, the weight of the difference is the integral:
\be
8 \pi \varepsilon \Omega^{\alpha \beta} \left[\int z \iota_{\pd_z} \bigg(P(z,t;0,-T) P(z,t;0,0)\bigg) + \bigg(T \to - T\bigg)\right]
\ee 
where $P(z,t;w,s)$ is the propagator of the $HT$ kinetic term which in the holomorphic gauge of \cite{GwilliamWilliams} is given by
\be
P(z,t;w,s) = \frac{(t-s)(\diff \oz - \diff \ow) -(\oz - \ow)(\diff t - \diff s)}{8\pi i \big((t-s)^2+|z-w|^2\big)^{3/2}} \diff z .
\ee
The resulting integral
\be
\begin{aligned}
	\frac{-\varepsilon\Omega^{\alpha \beta}}{8\pi} \int\diff^2 z \diff t \bigg[T|z|^2 & \bigg(\big((t-T)^2 + |z|^2)^{-3/2}\big(t^2 + |z|^2)^{-3/2}\\ 
	+ & \big((t+T)^2 + |z|^2)^{-3/2}\big(t^2 + |z|^2)^{-3/2}\bigg)\bigg]\\
\end{aligned}
\ee 
can be computed explicitly, giving $\varepsilon \Omega^{\alpha \beta}$ (since the diagrams are tree-level from the perspective of the bulk, there is no need to regularize). 
As expected, we find a commutator
\be
[Z^\alpha, Z^\beta] = \varepsilon \Omega^{\alpha \beta}
\ee
corresponding to the Weyl algebra associated to the symplectic form $\varepsilon\Omega^{\alpha \beta}$, cf. Section 3 of \cite{Yagi} or Section 6 of \cite{descent}.

\section{Superconformal Chern-Simons-Matter Theories}
\label{sec:CSM}
Although somewhat exotic, there are highly supersymmetric Chern-Simons-matter theories. Chern-Simons theories with $\CN=3$ supersymmetry go back to the works \cite{ZupnikKhetselius, KaoLee} and it was believed that this was the maximal $\CN=3$ supersymmetry attainable in the presence of Chern-Simons gauge fields, see e.g. \cite{KaoLeeLee, KSmirrorsym, GaiottoYin}. Somewhat surprisingly, there is a mechanism, first discovered by Gaiotto-Witten \cite{GaiottoWitten-Janus}, that allows for an $\CN=4$ coupling of hypermultiplets and Chern-Simons gauge fields; the resulting theories have since been known as Gaiotto-Witten theories. The work \cite{HLLLP1} generalized the Gaiotto-Witten theories to include both hypermultiplets (taking values in a representation $\CR$) and twisted hypermultiplets (taking values in a representation $\wt{\CR}$). The resulting theories have at least $\CN=4$ supersymmetry by the same mechanism found by Gaiotto-Witten, but include theories with yet more supersymmetry \cite{HLLLP2} such as the $\CN \geq 6$ ABJ(M) theories \cite{ABJM, ABJ} and $\CN=8$ BLG theory \cite{BL1, BL2, Gustavsson}.

In this section we describe the additional superconformal symmetries enjoyed in each of these cases, starting from $\CN=3$ and progressing to $\CN=8$. Representation theoretic aspects of these supersymmetry enhancements, and superpotentials arising below, were described in detail in \cite{dMFOME}.

\subsection{$\CN=3$ Chern-Simons-matter theories}
\label{sec:N=3CSM}

We start with the $\CN=3$ Chern-Simons-matter theories. This is a theory of an $\CN=2$ Chern-Simons vector multiplet $(\bA, \bB)$ coupled to an $\CN=4$ hypermultiplets $(\bZ, \bPsi)$ valued in a complex-symplectic representation $\CR$ of the complexified gauge group $G$. In addition, there is a superpotential coupling
\be
	W = \tfrac{\pi}{k}\Tr(\nu^2)\,.
\ee
We could also coupled to $\CN=4$ twisted hypermultiplets, but these two multiplets get identified upon reduction to $\CN=3$. Putting this information together, we arrive at the following $HT$-twisted action
\be
\label{eq:twistedactionN=3}
	S = \int \bB F'(\bA) + \tfrac{k}{4\pi} \Tr(\bA \pd \bA) + \bPsi \diff'_{\bA} \bZ + \tfrac{\pi}{k} \Tr(\bnu^2)
\ee
It follows that the action of the $HT$ supercharge takes the form
\be
\begin{aligned}
	Q_{HT} \bA & = F'(\bA) \qquad & Q_{HT} \bB & = \diff'_\bA \bB - \bmu + \tfrac{k}{2\pi}\pd \bA\\
	Q_{HT} \bZ & = \diff'_\bA \bZ \qquad & Q_{HT} \bPsi & = \diff'_\bA \bPsi + \tfrac{2\pi}{k} \Omega(-, \bnu \bZ)\\
\end{aligned}
\ee
where $\bnu \bZ$ denotes the action of $\bnu$ (viewed as an element of $\mathfrak{g}$ with the bilinear form $\Tr$) on $\bZ$. We give the hypermultiplets $R$-charge $\frac{1}{2}$, so that the (modified) stress tensor takes the form
\be
	\bT_V = \iota_V \bigg(-\bB \pd \bA + \tfrac{1}{4}\big(3 \bPsi \pd \bZ - \bZ \pd \bPsi\big)\bigg)
\ee

With a minor modification, an analog of the bosonic currents found in Section \ref{sec:SYM} can be found in this $\CN=3$ theory:
\be
	\bTheta_{\Gamma} = \tfrac{1}{2}\Gamma \big(\Omega^{-1}(\bPsi, \bPsi) - \Omega(\bZ, \pd_{\bB} \bZ)\big)
\ee
where $\pd_{\bB}$ denotes the holomorphic ``covariant derivative'' ${\pd_{\bB} = \pd - \tfrac{2\pi}{k} \bB}$. The fact that this is $Q_{HT}$-closed relies on the precise form of the superpotential $W$. The action of $\bTheta_{\Gamma}$ on the fields is given by
\be
\begin{aligned}
	\Gamma \cdot \bA & = \tfrac{2\pi}{k} \Gamma \bnu \qquad & \Gamma \cdot \bB & =0\\
	\Gamma \cdot \bZ & = \Gamma \Omega^{-1}(\bPsi, -) \qquad & \Gamma \cdot \bPsi & = -\Gamma \Omega(\pd_{\bB}\bZ, -) - \tfrac{1}{2}(\pd \Gamma) \Omega(\bZ,-)\\
\end{aligned}
\ee
This current transforms as a section of $K^{-1/2}_\C$ just as before:
\be
V\pd_z \cdot \textstyle{\int} \bTheta_{\Gamma} = \textstyle{\int}\bTheta_{-V \pd_z \Gamma + \scriptsize{\frac{1}{2}} (\pd_z V)\Gamma}
\ee

One immediate observation is that the action of $\Gamma$ on itself is not quite the above stress tensor:
\be
	\Gamma \cdot \textstyle{\int} \bTheta_{\bGamma'} = \textstyle{\int}\widehat{\bT}_{-2\Gamma \Gamma'}
\ee
where the ``stress tensor'' $\widehat{\bT}_V$ is given by
\be
	\widehat{\bT}_V = \iota_V \big(\tfrac{3}{4}\bPsi \pd_{\bB} \bZ - \tfrac{1}{4}\bZ \pd_{\bB} \bPsi\big)
\ee
The action of this current is given as follows:
\be
\begin{aligned}
	\widehat{V} \cdot \bA & = \tfrac{2\pi}{k} \iota_V\bnu \qquad & \widehat{V} \cdot \bB & = 0\\
	\widehat{V} \cdot \bZ & = V \pd_{\bB,z} \bZ + \tfrac{1}{4} (\pd_z V) \bZ \qquad & \widehat{V} \cdot \bPsi & = V \pd_{\bB,z} \bPsi + \tfrac{3}{4} (\pd_z V) \bPsi\\
\end{aligned}
\ee
The current $\widehat{\bT}_V$ \emph{still} generates an action of $\textrm{Vect}(\C)$ and, moreover, the two actions of $\textrm{Vect}(\C)$ are cohomologous:
\be
\begin{aligned}
	Q\bigg(\iota_V \big(\tfrac{\pi}{k} \Tr(\bB^2)\big)\bigg) & = \diff'\bigg(\iota_V \big(\tfrac{\pi}{k} \Tr(\bB^2)\big)\bigg) + \widehat{\bT}_V - \bT_V\\
	& \hspace{-2cm} \Rightarrow \{\textstyle{\int} \bT_V, - \}_{BV} = \{\textstyle{\int} \widehat{\bT}_V, - \}_{BV} + Q\big(...\big)
\end{aligned}
\ee

We see that the action of holomorphic vector fields $\textrm{Vect}(\C)$ gets enhanced to the Lie superalgebra
\be
	\lie{a}_{\CN=3} = \textrm{Vect}(\C) \oplus \Pi K^{-1/2}_\C
\ee
where we have written the Lie algebra as a module for the even part. If $\Gamma, \Gamma'$ are element of $K^{-1/2}_\C$, their bracket is given by
\be
[\Gamma , \Gamma'] = 2\Gamma' \Gamma.
\ee
Putting this together, we come to the following proposition.
\begin{proposition}
	The $HT$ twist of an $\CN=3$ superconformal Chern-Simons-matter theory admits an action of $\lie{a}_{\CN=3}$.
\end{proposition}

Notice that one can identity $\lie{a}_{\CN=3}$ with the positive part of the $\CN=1$ Virasoro algebra
\be
	\lie{a}_{\CN=3} = \lie{vir}_{\CN=1}^{\geq 0} .
\ee
\subsection{Enhancement to $\CN=4$}
\label{sec:N=4CSM}

As discovered by Gaiotto and Witten \cite{GaiottoWitten-Janus}, the above $\CN=3$ superconformal Chern-Simons theories have explicit supersymmetry enhancement when the representation $\CR$ satisfies the so-called fundamental identity: the theory has $\CN=4$ supersymmetry when the (complex) moment map for the $G$-action is isotropic with respect to the bilinear form $\Tr$, i.e. $\Tr(\nu^2) = 0$.%
\footnote{More precisely, the fundamental identity requires $\Tr(\nu^2)$ is constant. If we further require an unbroken $R$-symmetry, this further imposes that this constant vanishes.} %
In particular, the above superpotential vanishes identically. In terms of the representation matrices $(\tau_a)_{mn}$ and the matrix elements $K_{ab}$ and $K^{ab}$ of the pairing $\Tr$ and its dual, the fundamental identity says
\be
	K^{ab} \big((\tau_a)_{lm} (\tau_b)_{no} + (\tau_a)_{nm} (\tau_b)_{lo}\big) = 0
\ee
In the language of \cite{dMFOME}, the representation $\CR$ (together with the metric Lie algebra $(\fg, \Tr)$) forms an anti-Lie triple system (aLTS).

There is a somewhat larger class of $\CN=4$ Chern-Simons-matter theories, first described in \cite{HLLLP1}, where we couple to both hypermultiplets and twisted hypermultiplets; the latter will be valued in a (possibly different) symplectic representation $\wt{\CR}$ also satisfying the fundamental identity. Although the hypermultiplets (and similarly twisted hypermultiplets) are only coupled to one another via the gauge fields, there is a superpotential that couples the hypermultipets to the twisted hypermultiplets
\be
W = \tfrac{2\pi}{k}\Tr(\nu \wt{\nu})\,.
\ee
Putting this information together, we arrive at the following $HT$-twisted action for this generalized Gaiotto-Witten theory:
\be
\label{eq:twistedactionGGW}
	S = \int \bB F'(\bA) + \tfrac{k}{4\pi} \Tr(\bA \pd \bA) + \bPsi \diff'_{\bA} \bZ + \wt{\bPsi} \diff'_{\bA} \wt{\bZ} + \tfrac{ 2\pi}{k} \Tr(\bnu \wt{\bnu})
\ee
We give the hypermultiplets and twisted hypermultiplets $R$-charge $\frac{1}{2}$, and so the (modified) stress tensor takes the form
\be
	\bT_V = \iota_V \bigg(-\bB \pd \bA + \tfrac{1}{4}\big(3 \bPsi \pd \bZ - \bZ \pd \bPsi\big) + \tfrac{1}{4}\big(3 \wt{\bPsi} \pd \wt{\bZ} - \wt{\bZ} \pd \wt{\bPsi}\big)\bigg)
\ee
We note that the moment map for the $G$ action on $\CR \oplus \wt{\CR}$, i.e. $\nu + \wt{\nu}$, does not satisfy the fundamental identity and instead
\be
	\Tr\big((\nu+\wt{\nu})^2\big) = 2\Tr(\nu\wt{\nu})
\ee
In particular, this theory of hypermultipliets and twisted hypermultiplets is a special case of the above $\CN=3$ theory.

We now describe the currents generating an action of $\lie{a}_{\CN=4}$ commensurate with the fact that the underlying theory has $\CN=4$ supersymmetry. As in the Yang-Mills theories described in Section \ref{sec:SYM}, the hypermultiplets $\bZ$ have weight $1$ under the remnant $R$-symmetry, while the twisted hypermultiplets have weight $-1$. Thus, the conserved current associated to (the holomorphic extension of) this symmetry is simply
\be
	\bJ_S = S(z) \big(\bPsi \bZ - \wt{\bPsi} \wt{\bZ}\big)
\ee
The fundamental identity allows us to realize the bosonic current $\Theta_{\Gamma}$ as the sum of two currents:
\be
	\bTheta_{\Gamma} = \tfrac{1}{2}\Gamma \big(\Omega^{-1}(\bPsi, \bPsi) - \wt{\Omega}(\wt{\bZ}, \pd_{\bB} \wt{\bZ})\big) 
\ee
and
\be
	\wt{\bTheta}_{\wt{\Gamma}} = \tfrac{1}{2}\wt{\Gamma}\big(\wt{\Omega}^{-1}(\wt{\bPsi}, \wt{\bPsi}) - \Omega(\bZ, \pd_{\bB} \bZ)\big)
\ee
The action of $\bTheta_{\Gamma}$ on the fields is given by
\be
\begin{aligned}
	\Gamma \cdot \bA & = \tfrac{2\pi}{k} \Gamma \wt{\bnu} \qquad & \Gamma \cdot \bB & =0\\
	\Gamma \cdot \bZ & = \Gamma \Omega^{-1}(\bPsi, -) \qquad & \Gamma \cdot \bPsi & = 0\\
	\Gamma \cdot \wt{\bZ} & = 0 \qquad & \Gamma \cdot \wt{\bPsi} & = -\Gamma \wt{\Omega}(\pd_{\bB}\wt{\bZ}, -) - \tfrac{1}{2}(\pd \Gamma) \wt{\Omega}(\wt{\bZ},-)\\
\end{aligned}
\ee
The action of $\wt{\bTheta}_{\wt{\Gamma}}$ is nearly identical and given by exchanging hypermultiplets and twisted hypermultiplets.

As in the $\CN=3$ case, the action of $\Gamma$ on $\int \wt{\bTheta}_{\wt{\Gamma}}$ do not quite land on the stress tensor $\bT_V$:
\be
	\Gamma \cdot \textstyle{\int} \wt{\bTheta}_{\bGamma} = \textstyle{\int}\widehat{\bT}_{-\Gamma \wt{\Gamma}} + \textstyle{\int}\bJ_{\scriptsize{\frac{1}{4}}(\wt{\Gamma} \pd \Gamma - \Gamma \pd \wt{\Gamma})}
\ee
The ``stress tensor'' $\widehat{\bT}_V$ is given by
\be
	\widehat{\bT}_V = \iota_V \big(\tfrac{3}{4}(\bPsi \pd_{\bB} \bZ + \wt{\bPsi} \pd_{\bB} \wt{\bZ}) - \tfrac{1}{4}(\bZ \pd_{\bB} \bPsi + \wt{\bZ} \pd_{\bB} \wt{\bPsi})\big)
\ee
For the same reason as in the $\CN=3$ theories, the current $\widehat{\bT}_V$ \emph{still} generates an action of $\textrm{Vect}(\C)$ and this action is cohomologous to the one generated by $\bT_V$.

\begin{proposition}
	The $HT$ twist of any generalized Gaiotto-Witten theory admits an action of $\lie{a}_{\CN=4}$.
\end{proposition}

\subsection{Enhancement to $\CN=5$}
\label{sec:N=5CSM}

As pointed out in \cite{HLLLP2}, the $\CN=4$ supersymmetry in the general case enhances to $\CN=5$ supersymmetry if we ask that the hypermultiplets and twisted hypermultiplets transform in the same representation $\CR = \wt{\CR}$. In this case, we can write the hypermuletiplets and twisted hypermultiplets as a double $\bZ^\alpha =  (\bZ, \wt{\bZ})$. We can then make a triplet of complex moment map operators $\bnu^{\alpha \beta}_a = \frac12 (\tau_a)_{nm} \bZ^{\alpha n} \bZ^{\beta m} = \bnu^{\beta \alpha}_a$; the fundamental identity then implies
\be
	\Tr (\nu^{++}{}^2) = 0 = \Tr (\nu^{--}{}^2) \qquad \Tr (\nu^{++} \nu^{+-}) = 0 = \Tr (\nu^{--} \nu^{+-})
\ee
as well as
\be
	\Tr (\nu^{++} \nu^{--}) = - \Tr (\nu^{+-}{}^2)
\ee
With this notation, the superpotential can be written in a manifestly $\fsl(2)$-invariant fashion:
\be
	W = \tfrac{\pi}{2k} \epsilon_{\alpha \gamma} \epsilon_{\beta \delta} \Tr (\nu^{\alpha \beta} \nu^{\gamma \delta})
\ee
Correspondingly, the above $\fgl(1)$ symmetry gets extended by two additional currents
\be
	\bJ_{S_+} = S_+ \wt{\bPsi} \bZ \qquad \bJ_{S_-} = S_- \bPsi \wt{\bZ}
\ee
realizing this $\fsl(2)$ symmetry. We can compactly write these $\fsl(2)$ currents as
\be
	\bJ_S = S^i \bPsi_\alpha (\sigma_i)^\alpha{}_\beta \bZ^\beta
\ee
where $\sigma_i$ are a basis of $\mathfrak{sl}(2)$, e.g. the Pauli matrices. It is important to note that this $\mathfrak{sl}(2)$ symmetry is \emph{not} an $\CN=4$ flavor symmetry --- both currents $\bTheta_{\Gamma}, \wt{\bTheta}_{\wt{\Gamma}}$ transform non-trivially.%
\footnote{Recall that $\bTheta_{\Gamma}$ (resp. $\wt{\bTheta}_{\wt{\Gamma}})$ would be invariant under an $\CN=4$ Higgs (resp. Coulomb) flavor symmetry, cf. Section \ref{sec:N=4flavor}. Here, both $\bTheta_{\Gamma}$ and $\wt{\bTheta}_{\wt{\Gamma}}$ transform non-trivially under $\fsl(2)$. Indeed, they transform into one another!} %
Rather, these symmetries extend the remnant $\C^\times$ $R$-symmetry to $SL(2) \cong \Spin(3)$! (The original $\C^\times$ is realized as the diagonal torus.) Similarly, the two bosonic currents present in the $\CN=4$ setting gain a third so that the three transform in the adjoint representation of this $\fsl(2)$:
\be
	\bTheta_\Gamma = \tfrac{1}{2}\Gamma^{\alpha \beta} \bigg(\Omega^{-1}(\bPsi_\alpha, \bPsi_\beta) - \epsilon_{\alpha \gamma} \epsilon_{\beta \delta}\Omega(\bZ^\gamma, \pd_{\bB}\bZ^\delta)\bigg)
\ee
where $\Gamma^{\alpha \beta} = \Gamma^{\beta \alpha}$.

The symmetry algebra $\lie{a}_{\CN=5}$ realized by these currents takes the following form. Written as a module for its bosonic part, we have
\be
	\lie{a}_{\CN=5} = \textrm{Vect}(\C) \oplus \CO_\C \otimes \fsl(2) \oplus \Pi K^{-1/2}_\C \otimes \fsl(2)
\ee
If $\Gamma, \Gamma'$ are elements of $K^{-1/2}_\C \otimes \fsl(2)$, their bracket is given by
\be
	[\Gamma , \Gamma'] = \Tr(\Gamma' \Gamma^\vee) - \tfrac{1}{4}(\Gamma \pd \Gamma'^{\vee} + \Gamma' \pd \Gamma^{\vee})_0.
\ee
where $(\Gamma^\vee)_{\alpha \beta} = \epsilon_{\alpha \gamma} \epsilon_{\beta \delta} \Gamma^{\gamma \delta}$, and $(\Gamma' \Gamma)^\alpha{}_\beta = \Gamma'^{\alpha \gamma} \Gamma^\vee_{\beta \gamma}$; $M_0$ denotes the traceless part of the matrix $M$. These brackets imply that $\lie{a}_{\CN=5}$ is simply the positive part of the $\CN=3$ Virasoro algebra
\be
	\lie{a}_{\CN=5} = \lie{vir}_{\CN=3}^{\geq 0}
\ee

\begin{proposition}
	The $HT$ twist of any generalized Gaiotto-Witten theory where the hypermultiplets and twisted hypermultiplets transform in the same representation $\CR$ admits an action of $\lie{a}_{\CN=5}$.
\end{proposition}

\subsection{Enhancement to $\CN=6$}
\label{sec:N=6CSM}

The $\CN=5$ supersymmetry of the theory further enhances to $\CN=6$ when the representation $\CR$ splits as $\CR = R \oplus R^* = T^*R$ with its natural symplectic structure. We can then split the hypermultiplet as $\bZ = (\bX, \bY)$, with $\bX$ valued in the representation $R$ and $\bY$ valued in the dual $R^*$, and similarly for the twisted hypermultiplets $\wt{\bZ} = (\wt{\bX}, \wt{\bY})$ and the accompanying fermions. Let $(\rho_a)^s{}_t$ be the representation matrices for the $G$ action on $R$; it follows that the moment map can be expressed as $\nu_a = Y \rho_a X$, and the fundamental identity takes the following form:
\be
	K^{ab}\big( (\rho_a)^s{}_t (\rho_b)^u{}_v + (\rho_a)^s{}_v (\rho_b)^u{}_t \big) = 0
\ee
In the language of \cite{dMFOME}, the representation $R$ (together with the metric Lie algebra) forms an anti-Jordan triple system (aJTS).

As $\bX, \wt{\bX}$ (and $\wt{\bY},\bY$) transform in the same representation, we will collect them as $\bX^\alpha = (\bX, \wt{\bX})$ and $\bY^{\dot{\alpha}} = (\bY, \wt{\bY})$; we denote the accompanying fermionic fields $\bPsi_{\bX, \alpha}$ and $\bPsi_{\bY, \dot{\alpha}}$. With these operators, we can write four moment map operators $\bnu_a^{\alpha \dot{\alpha}} = \bY^{\dot{\alpha}} \rho_a \bX^{\alpha}$. Using the split form of the fundamental identity, the superpotential can be re-expressed in a manifestly $\mathfrak{sl}(2)_+ \oplus \mathfrak{sl}(2)_- \cong \mathfrak{so}(4)$ invariant fashion, where $\mathfrak{sl}(2)_+$ (resp. $\mathfrak{sl}(2)_-$) acts on the doublet $\bX^\alpha$ (resp. $\bY^{\dot{\alpha}}$):
\be
	W = \tfrac{\pi}{2k} \epsilon_{\alpha \beta} \epsilon_{\dot{\alpha} \dot{\beta}} \Tr(\nu^{\alpha \dot{\alpha}} \nu^{\beta \dot{\beta}})
\ee
The currents generating the action of $\CO_\C \otimes (\mathfrak{sl}(2)_+ \oplus \mathfrak{sl}(2)_-)$ are given by
\be
	\bJ_{S} = S^i \bPsi_{\bX, \alpha} (\sigma_i)^{\alpha}{}_{\beta} \bX^{\beta} \qquad \qquad \wt{\bJ}_{\wt{S}} = \wt{S}^i \bPsi_{\bY, \dot{\alpha}} (\sigma_i)^{\dot{\alpha}}{}_{\dot{\beta}} \bY^{\dot{\beta}} 
\ee
The supercharges $\bTheta_{\Gamma}$ transforming in $K^{-1/2}_\C$ gain a fourth partner to form a $\Spin(4)$ vector
\be
	\bTheta_{\Gamma} = \Gamma^{\alpha \dot{\alpha}}\bigg(\bPsi_{\bX, \alpha} \bPsi_{\bY, \dot{\alpha}} - \tfrac{1}{2} \epsilon_{\alpha \beta} \epsilon_{\dot{\alpha} \dot{\beta}} \big(\bX^\beta \pd_{\bB} \bY^{\dot{\beta}} - \bY^{\dot{\beta}} \pd_{\bB} \bX^\beta \big) \bigg)
\ee

One new feature is that there is a non-trivial mixing between the action of $\mathfrak{so}(4)$ and $\CO_\C$. Namely, we find that the action of $\CO_\C \otimes \mathfrak{sl}(2)_\pm$ is given by
\be
\begin{aligned}
	S \cdot \textstyle{\int}\bTheta_{\Gamma} = \textstyle{\int}\bTheta_{-(S\otimes \id) \Gamma} + \textstyle{\int}\bXi_{\scriptsize{\frac{1}{2}}\Gamma^\vee (\pd S \otimes \id)}\\
	 \wt{S} \cdot \textstyle{\int}\bTheta_{\Gamma} = \textstyle{\int}\bTheta_{-(\id \otimes \wt{S})\Gamma} + \textstyle{\int}\bXi_{-\scriptsize{\frac{1}{2}} \Gamma^\vee (\id \otimes \pd \wt{S})}\\
\end{aligned}
\ee
where the additional $\Spin(4)$ (co)vector of $K^{1/2}_\C$-valued currents $\bXi_{\Sigma}$ is given by 
\be
\bXi_\Sigma = \bSigma_{\alpha \dot{\alpha}} \bY^{\dot{\alpha}} \bX^\alpha
\ee
and $\Gamma^\vee_{\alpha \dot{\alpha}} = \epsilon_{\alpha \beta} \epsilon_{\dot{\alpha} \dot{\beta}} \Gamma^{\beta \dot{\beta}}$ is the bi-spinor avatar of lowering a vector index with the Euclidean metric on $\C^4$. In particular, these currents transform under $\textrm{Vect}(\C) \oplus \CO_\C \otimes \mathfrak{so}(4)$ as an extension of $K^{-1/2}_\C \otimes (\C^2_+ \times \C^2_-)$ by $K^{1/2}_\C \otimes (\C^2_+ \times \C^2_-)^*$.

We note that there is an additional $\fgl(1)$ symmetry coming from the splitting of $\CR$:
\be
	\bL_{U} = U(z)(\bPsi_{\bX, \alpha} \bX^\alpha - \bPsi_{\bY, \dot{\alpha}} \bY^{\dot{\alpha}})
\ee
The corresponding $\CO_\C \otimes \fgl(1)$ symmetry commutes with the $\CO_\C \otimes \mathfrak{so}(4)$ $R$-symmetry and acts trivially on the $\bXi_{\Sigma}$, but has a non-trivial action on $\bTheta_{\Gamma}$:
\be
U \cdot \textstyle{\int}\bTheta_{\Gamma} = \textstyle{\int} \bXi_{\Gamma^\vee \pd U}
\ee
In particular, this implies that $\bL_{U}$ centrally extends the $\CO_\C \otimes \mathfrak{so}(4)$ remnant $R$-symmetry.%
\footnote{If this were a flavor symmetry then the action of $\bL_{U}$ on $\bTheta_{\Gamma}$ would yield some new current rather than $\bXi_\Sigma$.} %

Putting this together, we have a Lie superalgebra $\lie{a}_{\CN=6}$ whose bosonic subalgebra is
\be
	(\lie{a}_{\CN=6})^+ = \textrm{Vect}(\C) \oplus \CO_\C \otimes \big(\mathfrak{so}(4) \oplus \fgl(1)\big)
\ee
and with fermionic subspace given by the extension
\be
	0 \to K^{1/2}_\C \otimes (\C^2_+ \otimes \C^2_-)^*_0 \to (\lie{a}_{\CN=6})^- \to K^{-1/2}_\C \otimes (\C^2_+ \otimes \C^2_-)_0 \to 0 .
\ee
In this expression, we use the subscript to denote the $\fgl(1)$ weight. Let $\Gamma$ denote an element of $K^{-1/2}_\C \otimes (\C^2_+ \otimes \C^2_-)_0$ and let $\Sigma$ an element of $K^{1/2}_\C \otimes (\C^2_+ \otimes \C^2_-)^*_0$.
The brackets between odd elements takes the form: $[\Sigma, \Sigma'] = 0$, 
\be
	[\Sigma, \Gamma] =  (\Gamma \Sigma^T)_0 -(\Gamma^T \Sigma)_0 + \Tr(\Sigma \Gamma^T) 
\ee
and
\be
\begin{split}
	[\Gamma , \Gamma'] &= \Tr(\Gamma'^T \Gamma^\vee) - \tfrac{1}{4}(\Gamma \pd \Gamma'^{\vee T} + \Gamma' \pd \Gamma^{\vee T})_0 \\
	&-\tfrac{1}{4}(\Gamma^T \pd \Gamma'^{\vee} + \Gamma'^T \pd \Gamma^{\vee})_0 - \tfrac{1}{4}\Tr(\Gamma \pd \Gamma'^{\vee T} + \Gamma' \pd \Gamma^{\vee T}).
\end{split}
\ee
We can thus identify $\lie{a}_{\CN=6}$ with the positive part of the (big) $\CN=4$ superconformal algebra $K'_4$
\be
	\lie{a}_{\CN=6} = K'_4{}^{\geq 0} .
\ee

\begin{proposition}
	The $HT$ twist of any generalized Gaiotto-Witten theory where the hypermultiplets and twisted hypermultiplets transform in the same split representation $R\oplus R^*$ admits an action of $\lie{a}_{\CN=6}$.
\end{proposition}

\subsection{Enhancement to $\CN=8$}
There is a further, explicit enhancement to $\CN=8$ supersymmetry when the representation $R$ possesses a $G$-invariant metric $g_{uv}$, so that $R$ is self-dual $R \cong R^*$ and $(\rho_a)_{uv} = g_{ut} (\rho_a)^{t}{}_v$ is antisymmetric in $u,v$, from which the fundamental identity implies that the tensor $K^{ab}(\rho_a)_{st}(\rho_b)_{uv}$ is totally antisymmetric in $s,t,u,v$. A non-trivial (i.e. non-free) example of this enhancement occurs when $\mathfrak{g} = \mathfrak{so}(4) \cong \mathfrak{sl}(2)_+ \oplus \mathfrak{sl}(2)_-$ and $R = \C^4 \cong \C^2_+ \oplus \C^2_-$, or perhaps decoupled copies thereof, whence $K^{ab}(\rho_a)_{st}(\rho_b)_{uv} \propto \epsilon_{stuv}$. The resulting theory was found independently by Bagger-Lambert \cite{BL1, BL2} and Gustovsson \cite{Gustavsson} and has since been called BLG theory.%
\footnote{There are ``higher rank'' versions of the BLG theory, discovered by Aharony-Bergman-Jaffries-Maldacena (ABJM) \cite{ABJM}, but the enhancement from $\CN=6$ to $\CN=8$ involves currents built from monopole operators.} %
In the language of \cite{dMFOME}, the representation $R$ is said to be the complexification of a representation giving a triple system of type 3-Lie algebra (3LA).

Using the above $G$-invariant metric, we can organize the hypermultiplet and twisted hypermultiplet scalar fields into four $R$-valued fields $\bZ^I = (\bX, \bY^\vee, \wt{\bX}, \wt{\bY}^\vee)$, and similarly organize the fermionic fields into four $R^*$-valued fields $\bPsi_I$. With this notation, we can express the superpotential as
\be
	W = \tfrac{\pi}{3! k} \epsilon_{IJKL} K^{ab} \rho_a(Z^I, Z^J)\rho_b(Z^K, Z^L)\,.
\ee
This expression makes manifest an $SL(4) \cong \Spin(6)$ invariance of the superpotential. The theory therefore admits an action of the bosonic Lie algebra
\be
(\lie{a}_{\CN=8})^+ = \textrm{Vect}(\C) \oplus \CO_\C \otimes \mathfrak{sl}(4)
\ee
The currents generating the $\CO_\C \otimes \mathfrak{sl}(4)$ action are given by
\be
\bL_{U} = U^{I}{}_J \bPsi_{I} \bZ^{J}
\ee 
where $U$ must to be traceless for this local operator to be $Q_{HT}$-closed. The current for the $\textrm{Vect}(\C)$ action is given as before.

Acting with $SL(4)$ on the $\CN=6$ fermionic symmetries, we arrive at the following currents:
\be
\bXi_{\Sigma} = \tfrac{1}{2} \Sigma_{IJ} \bZ^{I} \bZ^{J} \qquad \bTheta_{\Gamma} =  \tfrac{1}{2}\Gamma^{IJ} \bigg(\bPsi_{I} \bPsi_{J} - \tfrac{1}{2}\epsilon_{IJKL} \bZ^{K} \pd_\bB \bZ^{L}\bigg)
\ee
with $\Sigma_{IJ} = \Sigma_{JI}$ and $\Gamma^{IJ} = -\Gamma^{JI}$. We have left implicit the $G$-invariant metric on $R$. The action of $\CO_\C \otimes \mathfrak{sl}(4)$ on these fermionic symmetry generators takes the form
\be
\begin{aligned}
	U \cdot \textstyle{\int} \bXi_{\Sigma} & = \textstyle{\int} \bXi_{\Sigma(U \otimes \id + \id \otimes U)}\\
	U \cdot \textstyle{\int} \bTheta_{\Gamma} & = \textstyle{\int} \bTheta_{-(U \otimes \id + \id \otimes U)\Gamma} + \textstyle{\int} \bXi_{-(\star \Gamma) (\pd U \otimes \id + \id \otimes \pd U)}\\
\end{aligned}
\ee
where $(\star\Gamma)_{IJ} = \tfrac{1}{2} \epsilon_{IJKL} \Gamma^{KL}$. Together with the action of holomorphic vector fields $\textrm{Vect}(\C)$, we find that these odd symmetries transform as the extension
\be
0 \to K^{1/2}_\C \otimes \Sym^2(\C^4)^* \to (\lie{a}_{\CN=8})^- \to K^{-1/2}_\C \otimes \textstyle{\bigwedge}^2 \C^4 \to 0
\ee
Just as in the $\CN=6$ case, the bracket between two elements of $K^{1/2}_\C \otimes \Sym^2(\C^4)^*$ always vanishes. The bracket between $K^{1/2}_\C \otimes \Sym^2(\C^4)^*$ and $K^{-1/2}_\C \otimes \bigwedge^2\C^4$ is simply
\be
\Sigma \cdot \textstyle{\int} \bTheta_\Gamma = \textstyle{\int} \bL_{-\Sigma \Gamma}
\ee
and the bracket of two elements of $K^{-1/2}_\C \otimes \bigwedge^2\C^4$ takes the form
\be
\Gamma \cdot \textstyle{\int} \bTheta_{\Gamma'} = \textstyle{\int} \bT_{\scriptsize{\frac{1}{2}} \Tr(\Gamma' \star \Gamma)} + \textstyle{\int} \bL_{\scriptsize{\frac{1}{2}}(\Gamma' \star \pd \Gamma + \Gamma \star \pd \Gamma')}
\ee

The brackets described above are precisely those of the exceptional simple Lie superalgebra which is denoted
\beqn
\lie{a}_{\cN=8} = E(1|6)
\eeqn
in \cite{KacClass}.
Thus we obtain the following.

\begin{proposition}
	The $HT$ twist of BLG theory admits an action of the exceptional simple Lie superalgebra~$E(1|6)$.
\end{proposition}

Strictly speaking, the version of $E(1|6)$ we find in the proposition above is an analytic version of the one studied in \cite{KacClass} which is constructed using formal power series.
At the level of local operators one can show is that the true version of $E(1|6)$, defined using formal power series, acts.

\subsubsection{Non-perturbative $\CN=8$ Enhancement of rank 1 ABJM}
\label{sec:blgl1}

In this subsection we consider another example with $\CN=8$ supersymmetry, the rank 1 ABJM theory modeling the worldvolume of a single M2 brane probing $\C^4/\Z_k$. We illustrate how the symmetry enhancement works for $k =1, 2$.

An $E(1|6)$ symmetry of the minimally twisted rank 1 ABJM theory at level $k=1$, was identified in \cite{twistedgraviton} from the perspective of a putatively holographically dual twisted supergravity theory. There, we studied the minimal twist of eleven-dimensional supergravity on $AdS_4\times S^7$. The theory has a $U(1)$ global symmetry that mixes rescalings normal to the conformal boundary of $AdS_4$ with a global rescaling of the space of fields. This symmetry induces a consistent grading on the space of fields of depth $-2$, and has the following features:
\begin{itemize}
	\item The weight zero component is a certain local Lie algebra on $\R\times \C$ whose $\infty$-jets at the origin is $E(1|6)$.  
	\item The nonzero weight components carry actions of the weight zero component.
	\item The weight $-1$ component describes the holomorphic-topological twist of the rank 1 ABJM theory at level $k=1$. 
\end{itemize}

As in the previous sections of this paper, here we will describe the action of $\lie{a}_{\CN=8}$ in terms of explicit currents, which render the aforementioned $E(1|6)$ symmetry inner. However, the $\CN=8$ superconformal symmetry of the ABJM theory is not visible at the level of the Lagrangian and is due to nonperturbatives effects. Accordingly, the symmetry enhancement enhancement is non-perturbative in nature: the currents realizing $\lie{a}_{\CN=8}$ will include monopole operators. 

Monopole operators in $HT$ twisted $\CN=2$ Chern-Simons gauge theories are in general quite difficult to study; in contrast, work of Zeng \cite{Zeng} provides an explicit description of the DG vector space of local operators using the state-operator correspondence together with the explicit geometric quantization of the equations of motion on the raviolo. We will apply this analysis to the rank 1 ABJM theory and illustrate the enhancement to $\CN=8$ for $k = 1, 2$. For levels $k > 2$, such an enhancement is not present, and though the above prescription yields the correct space of local operators, and new techniques are needed to compute the algebra structure. 

As a reminder, the rank 1 ABJM theory is a $U(1)_k \times U(1)_{-k}$ Chern-Simons gauge theory with chiral multiplets $X^\alpha$ of weight $(1,-1)$ and $Y^{\dot{\alpha}}$ of weight $(-1,1)$; written as an $\CN=2$ theory, there is no superpotential. Without loss, we take $k > 0$. The $HT$ twisted action then takes the form
\be
\label{eq:actionBLGL1}
\begin{aligned}
	S & = \int \bB_+ \diff' \bA_+ + \tfrac{k}{4\pi}\bA_+ \pd \bA_+ + \bB_- \diff' \bA_- - \tfrac{k}{4\pi}\bA_- \pd \bA_-\\
	& \qquad + \int \bPsi_{\bX,\alpha} \diff'_\bA \bX^\alpha + \bPsi_{\bY, \dot{\alpha}} \diff'_\bA \bY^{\dot{\alpha}}
\end{aligned}
\ee
with corresponding action of the $HT$ supercharge
\be
\label{eq:QBLGL1}
\begin{aligned}
	Q_{HT} \bA_\pm & = \diff' \bA_\pm \qquad & Q_{HT} \bB_\pm & = \diff'\bB_\pm \mp \bmu \pm \tfrac{k}{2\pi}\pd \bA_\pm\\
	Q_{HT} \bX^\alpha & = \diff'_\bA \bX^\alpha \qquad & Q_{HT} \bPsi_{\bX, \alpha} & = \diff'_\bA \bPsi_{\bX, \alpha}\\
	Q_{HT} \bY^{\dot{\alpha}} & = \diff'_\bA \bY^{\dot{\alpha}} \qquad & Q_{HT} \bPsi_{\bY,\dot{\alpha}} & = \diff'_\bA \bPsi_{\bY,\dot{\alpha}}\\
\end{aligned}
\ee
In this expression, $\bmu = \bPsi_{\bX, \alpha} \bX^\alpha - \bPsi_{\bY, \dot{\alpha}} \bY^{\dot{\alpha}}$ is the moment map for the $U(1)_k$ action on the parity shifted cotangent bundle or, equivalently, the negative of the moment map for $U(1)_{-k}$.

Monopole operators are local operators that source gauge fields $\bA_\pm$ that are partial connections on non-trivial holomorphic $\C^\times$ bundles on the radial $\P^1$. In particular, the moduli of such bundles has disconnected components labeled by $\Z^2$, the magnetic charge of the monopole operator, or better the ``monopole number.'' Note that the Chern-Simons terms induce an electric charge on magnetically charged objects: a bare monopole of magnetic charge $(\mathfrak{m}_+, \mathfrak{m}_-) \in \Z^2$ will have induced electric charge $(k \mathfrak{m}_+, -k \mathfrak{m}_-)$. This implies there can be no gauge-invariant monopole operators for magnetic charges with $\mathfrak{m}_+ - \mathfrak{m}_- \neq 0$ because the only charged fields have weights $(q_+, q_-)$ with $q_+ = - q_-$. We will denote the bare monopole operator of magnetic charge $(\mathfrak{m},\mathfrak{m})$ by $\bV_\mathfrak{m}$; this can be identified as $\bV_\mathfrak{m} = e^{2\pi \mathfrak{m} \bGamma}$, where $\bGamma$ is the corresponding dual photon and satisfies
\be
\pd \bGamma = \bB_+ + \bB_-	\qquad Q \bV_{\mathfrak{m}} = \diff' \bV_{\mathfrak{m}} + k \mathfrak{m}(\bA_+ - \bA_-) \bV_{\mathfrak{m}}
\ee
cf. \cite[Section 3.1]{CostelloDimofteGaiotto-boundary}. With the $\bV_{\mathfrak{m}}$, we can realize gauge-invariant local operators as $Q$-cohomology classes of $U(1) \times U(1)$-invariants built from the above fields, minus the zeromodes of $\bA_\pm$.

At monopole number zero, there are the familiar $\CN = 6$ currents $\bL_U, \wt{\bL}_{\wt{U}}$ and $\bTheta_\Gamma, \wt{\bTheta}_{\wt{\Gamma}}$ as well as the additional central currents $\bL_T = T(z)(\bPsi_{\bX, \alpha} \bX^\alpha)$ and $\wt{\bL}_{\wt{T}} = \wt{T}(z)(\bPsi_{\bY, \dot{\alpha}} \bY^{\dot{\alpha}})$. The difference $\bL_{T} - \wt{\bL}_T = T(z) \bmu$ is cohomologous to the current $\tfrac{1}{4 \pi} T(z) (\pd \bA_+ + \pd \bA_-)$ (up to a total derivative) generating the topological flavor symmetry measuring monopole number.

Now consider local operators of monopole number $\mathfrak{m} = \pm1$, i.e. a gauge-invariant local operator built by dressing the operators $\bV_{\pm 1}$. To get something gauge-invariant, the dressing factor must have electric charge $(\mp k, \pm k)$. As a module for the operators of monopole number 0, these are generated by degree $k$ polynomials in $\bX^\alpha, \bPsi_{\bX, \alpha}$ and their $\pd_z$ derivatives times $\bV_1$ for $\mathfrak{m}=1$, and similarly for $\mathfrak{m}=-1$. For $k = 1$, there are eight operators to consider:
\be
k = 1: \qquad \overset{\bM^I}{\overbrace{\bX^1 \bV_{-1}, \bY^{\dot{1}} \bV_{1}, \bX^2 \bV_{-1}, \bY^{\dot{2}} \bV_{-1}}}, \overset{\bDelta_I}{\overbrace{\bPsi_{\bX,1} \bV_{1}, \bPsi_{\bY, \dot{1}} \bV_{-1}, \bPsi_{\bX, 2} \bV_{1}, \bPsi_{\bY, \dot{2}} \bV_{-1}}}
\ee
It is straightforward to check that the above currents are simply various bilinears of these basic operators. The full set of currents generating the $\lie{a}_{\CN=8}$ action take the form as in the previous section, up to identifying $\bM^I \leftrightarrow \bZ^I$ and $\bDelta_I \leftrightarrow \bPsi_I$. Note that monopole number is identified with the weight under the diagonal matrix $\textrm{diag}(-1,1,-1,1)$ in $\fsl(4)$; the perturbative $\CN=6$ algebra is the subalgebra of weight $0$.

We also note the additional $\CN=8$ enhancement for $k = 2$: monopole number is identified with half of the above $\fsl(4)$ weight, with the monopole number $-1$ sector generated by the odd currents $\bPsi_{\bY, \dot{\alpha}} \bX^\alpha \bV_{-1}$ as well as the even currents $\bX^\alpha \bX^\beta \bV_{-1}$ and $(\bPsi_{\bY, \dot{1}} \bPsi_{\bY,\dot{2}} - \tfrac{1}{2}(\bX^1 \pd_\bB \bX^2 - \bX^2 \pd_\bB \bX^1))\bV_{-1}$. The additional generators at monopole number $1$ take the same form with $\bX \leftrightarrow \bY$ and $\bV_{-1} \to \bV_1$.

\subsubsection{$B$-type superconformal deformation}
From the perspective of the $\CN=4$ algebra, the topological $B$ twist corresponds to deforming the action by the element $\bTheta_{\Gamma}$ for $\Gamma^{34} = 1$, with the other components vanishing. It follows that the homotopy trivializing rotations in this $B$-twist is realized by $\bTheta_\Gamma$ for $\Gamma^{12} = z$. We can determine the result of turning on an $\Omega$ background by deforming the action $S$ by this superconformal element, which we denote by $\bTheta$. 

A straightforward, albeit it tedious, computation shows that the cohomology of $\lie{a}_{\CN=8}$ with respect to $\bTheta$ is as follows. The even cohomology can be identified with the bosonic subalgebra $\fsl(2)_{12} \oplus \fsl(2)_{34}$ rotating the $12$ and $34$ planes; the fermionic part of the cohomology can be identified with $\Pi \fsl(2)_{34}$, arising from the $K^{1/2}$ generators with $\Sigma_{ij} = 1$ for $I,J \in \{3,4\}$. From an $\CN=4$ perspective, $\fsl(2)_{12}$ (resp. $\fsl(2)_{34}$) should be thought of as a Higgs-branch (resp. Coulomb-branch) flavor symmetry; these cohomology classes can be identified with those found in Section \ref{sec:N=4flavor}: $\fsl(2)_{12}$ acts by rotating the Higgs branch and deforming by the elements of $\Pi \fsl(2)_{34}$ deforms the Higgs branch.

Let us return to the example of the rank 1 ABJM theory, focusing on $k = 1$ for simplicity. Local operators surviving the $\Omega$ deformation can be obtained via a spectral sequence: the first page computes the cohomology of $Q_{HT}$, which is generated by the 0-form components $M^I, \Delta_I$ of $\bM^I,\bDelta_I$ and their $\pd_z$ derivatives. The second page corresponds to turning on the differential induced by $\bTheta$: the action of $\bTheta$ on the fundamental fields is given by
\be
\begin{aligned}
	\big\{\textstyle{\int} \bTheta, \bA_\pm\big\} & = - 2 \pi z \wt{\bnu} \qquad & \big\{\textstyle{\int} \bTheta, \bB_\pm\big\} & = 0\\
	\big\{\textstyle{\int} \bTheta, \bZ^\alpha\big\} & = z \epsilon^{\alpha \beta} \bPsi_\beta \qquad & \big\{\textstyle{\int} \bTheta, \bPsi_\alpha\big\} & = 0\\
	\big\{\textstyle{\int} \bTheta, \bZ^{\dot{\alpha}}\} & = 0 \qquad & \big\{\textstyle{\int} \bTheta, \wt{\bPsi}_{\dot{\alpha}}\big\} & = -\epsilon_{\dot{\alpha} \dot{\beta}}(z \CD_z \wt{\bZ}^{\dot{\beta}} + \tfrac{1}{2}\wt{\bZ}^{\dot{\beta}})\\
\end{aligned}
\ee
where $\pd_\bB = \CD_z \diff z$. The fact that $\bTheta$ acts trivially on $\bB_\pm$ implies it will also act trivially on $\bGamma$ and hence the monopole operators $\bV_m$, thus the action of $\bTheta$ on the local operators $M^I$ and $\Delta_I$ is given by
\be
	\big\{\textstyle{\int} \bTheta, M^I\big\} = z\Delta_J (\delta^{I1}\delta^{J2} - \delta^{I2}\delta^{J1}) \qquad \big\{\textstyle{\int} \bTheta, \Delta_I\big\} = \big(z \pd_z M^J + \tfrac{1}{2} M^J\big)(\delta_{I3}\delta_{J4} - \delta_{I4}\delta_{J3})
\ee

As in Section \ref{sec:hyperSC}, the only local operators that survive the $\Omega$ deformation are generated by $M^1(0)$ and $M^2(0)$. We expect that the 1-shifted Poisson bracket on local operators transfers to a bracket between $M^1(0)$ and $M^2(0)$ on $\bTheta$-cohomology. 

\begin{conjecture}
The $\bTheta$-cohomology of local operators in the $HT$-twisted rank 1 level 1 ABJM theorem is the algebra of differential operators on $\C$. 
\end{conjecture}

Note that the algebra of differential operators on $\C$ is the same thing as the spherical Cherednik algebra for $\fgl(1)$ which is the expected answer for the quantized Coulomb branch algebra for the rank 1 level 1 ABJM theory. 


\bibliography{HTenhance}

\begin{thebibliography}{10}

\bibitem{SWsuco}
I.~Saberi and B.~R. Williams, ``Superconformal algebras and holomorphic field
  theories,'' {\em Ann. Henri Poincar\'{e}}, vol.~24, no.~2, pp.~541--604,
  2023.

\bibitem{GWrav}
N.~Garner and B.~R. Williams, ``{Raviolo vertex algebras},'' 8 2023.

\bibitem{GRW2}
N.~Garner, S.~Raghavendran, and B.~R. Williams, ``Higgs and coulomb branches
  from raviolo vertex algebras,'' 2023.

\bibitem{ZupnikKhetselius}
B.~M. Zupnik and D.~V. Khetselius, ``{Three-dimensional extended supersymmetry
  in the harmonic superspace. (In Russian)},'' {\em Sov. J. Nucl. Phys.},
  vol.~47, pp.~730--735, 1988.

\bibitem{KaoLee}
H.-C. Kao and K.-M. Lee, ``{Selfdual Chern-Simons systems with an N=3 extended
  supersymmetry},'' {\em Phys. Rev. D}, vol.~46, pp.~4691--4697, 1992.

\bibitem{GaiottoYin}
D.~Gaiotto and X.~Yin, ``Notes on superconformal chern-simons-matter
  theories,'' {\em JHEP}, vol.~2007, p.~056–056, Aug 2007.

\bibitem{KSmirrorsym}
A.~Kapustin and M.~J. Strassler, ``{On mirror symmetry in three-dimensional
  Abelian gauge theories},'' {\em JHEP}, vol.~04, p.~021, 1999.

\bibitem{dMFOME}
P.~de~Medeiros, J.~Figueroa-O'Farrill, and E.~Mendez-Escobar,
  ``{Superpotentials for superconformal Chern-Simons theories from
  representation theory},'' {\em J. Phys. A}, vol.~42, p.~485204, 2009.

\bibitem{GaiottoWitten-Janus}
D.~Gaiotto and E.~Witten, ``Janus configurations, chern-simons couplings, and
  the $\theta$-angle in $ \mathcal{N} = 4 $ super yang-mills theory,'' {\em
  JHEP}, vol.~2010, Jun 2010.

\bibitem{HLLLP1}
K.~Hosomichi, K.-M. Lee, S.~Lee, S.~Lee, and J.~Park, ``$\mathcal{N} = 4$
  superconformal chern-simons theories with hyper and twisted hyper
  multiplets,'' {\em JHEP}, vol.~2008, p.~091–091, Jul 2008.

\bibitem{HLLLP2}
K.~Hosomichi, K.-M. Lee, S.~Lee, S.~Lee, and J.~Park, ``{N=5,6 Superconformal
  Chern-Simons Theories and M2-branes on Orbifolds},'' {\em JHEP}, vol.~09,
  p.~002, 2008.

\bibitem{KacClass}
V.~G. Kac, ``Classification of infinite-dimensional simple linearly compact lie
  superalgebras,'' {\em Adv. Math.}, vol.~139, no.~1, pp.~1--55, 1998.

\bibitem{twistedgraviton}
S.~Raghavendran and B.~R. Williams, ``Twisted graviton spectra for ads4 and
  ads7,'' 2023.

\bibitem{BL1}
J.~Bagger and N.~Lambert, ``{Modeling Multiple M2's},'' {\em Phys. Rev. D},
  vol.~75, p.~045020, 2007.

\bibitem{BL2}
J.~Bagger and N.~Lambert, ``{Gauge symmetry and supersymmetry of multiple
  M2-branes},'' {\em Phys. Rev. D}, vol.~77, p.~065008, 2008.

\bibitem{Gustavsson}
A.~Gustavsson, ``{Algebraic structures on parallel M2-branes},'' {\em Nucl.
  Phys. B}, vol.~811, pp.~66--76, 2009.

\bibitem{ABJM}
O.~Aharony, O.~Bergman, D.~L. Jafferis, and J.~Maldacena, ``{N=6 superconformal
  Chern-Simons-matter theories, M2-branes and their gravity duals},'' {\em
  JHEP}, vol.~10, p.~091, 2008.

\bibitem{ABJ}
O.~Aharony, O.~Bergman, and D.~L. Jafferis, ``{Fractional M2-branes},'' {\em
  JHEP}, vol.~11, p.~043, 2008.

\bibitem{GRWthf}
O.~Gwilliam, E.~Rabinovich, and B.~R. Williams, ``{Quantization of
  topological-holomorphic field theories: local aspects},'' 7 2021.

\bibitem{AganagicCostelloMcNamaraVafa}
M.~Aganagic, K.~Costello, J.~McNamara, and C.~Vafa, ``Topological
  chern-simons/matter theories,'' 2017.

\bibitem{CostelloDimofteGaiotto-boundary}
K.~Costello, T.~Dimofte, and D.~Gaiotto, ``{Boundary Chiral Algebras and
  Holomorphic Twists},'' {\em Commun. Math. Phys.}, vol.~399, no.~2,
  pp.~1203--1290, 2023.

\bibitem{EagerSaberiWalcher}
R.~Eager, I.~Saberi, and J.~Walcher, ``Nilpotence varieties,'' {\em Ann. Henri
  Poincar{\'e}}, 2021.

\bibitem{EStwists}
C.~Elliott and P.~Safronov, ``Topological twists of supersymmetric algebras of
  observables,'' {\em Commun. Math. Phys.}, vol.~371, p.~727–786, Mar 2019.

\bibitem{AHISS}
O.~Aharony, A.~Hanany, K.~A. Intriligator, N.~Seiberg, and M.~J. Strassler,
  ``{Aspects of N=2 supersymmetric gauge theories in three-dimensions},'' {\em
  Nucl. Phys.}, vol.~B499, pp.~67--99, 1997.

\bibitem{BV}
I.~Batalin and G.~Vilkovisky, ``{Gauge Algebra and Quantization},'' {\em Phys.
  Lett. B}, vol.~102, pp.~27--31, 1981.

\bibitem{CostelloRenormalization}
K.~J. Costello, ``Renormalization and effective field theory,'' in {\em
  Mathematical Surverys and Monographs}, vol.~170, American Mathematical
  Society, 2011.

\bibitem{GwilliamWilliams}
O.~Gwilliam and B.~R. Williams, ``A one-loop exact quantization of chern-simons
  theory,'' 2019.

\bibitem{OhYagi}
J.~Oh and J.~Yagi, ``{Poisson vertex algebras in supersymmetric field
  theories},'' {\em Lett. Math. Phys.}, vol.~110, no.~8, pp.~2245--2275, 2020.

\bibitem{Zeng}
K.~Zeng, ``{Monopole operators and bulk-boundary relation in holomorphic
  topological theories},'' {\em SciPost Phys.}, vol.~14, no.~6, p.~153, 2023.

\bibitem{CG}
K.~Costello and O.~Gwilliam, {\em {Factorization Algebras in Quantum Field
  Theory. Vol. 1}}, vol.~31 of {\em New Mathematical Monographs}.
\newblock Cambridge University Press, 2016.

\bibitem{CG2}
K.~Costello and O.~Gwilliam, {\em Factorization {A}lgebras in {Q}uantum {F}ield
  {T}heory. {V}ol. 2}, vol.~41 of {\em New Mathematical Monographs}.
\newblock Cambridge University Press, 2021.

\bibitem{twistedN=4}
N.~Garner, ``{Twisted Formalism for 3d $\mathcal{N}=4$ Theories},'' 4 2022.

\bibitem{Yagi}
J.~Yagi, ``$\omega$-deformation and quantization,'' {\em JHEP}, vol.~2014, Aug
  2014.

\bibitem{descent}
C.~Beem, D.~Ben-Zvi, M.~Bullimore, T.~Dimofte, and A.~Neitzke, ``{Secondary
  products in supersymmetric field theory},'' {\em Ann. Henri Poincar{\'e}},
  vol.~21, no.~4, pp.~1235--1310, 2020.

\bibitem{KaoLeeLee}
H.-C. Kao, K.-M. Lee, and T.~Lee, ``{The Chern-Simons coefficient in
  supersymmetric Yang-Mills Chern-Simons theories},'' {\em Phys. Lett. B},
  vol.~373, pp.~94--99, 1996.

\end{thebibliography}
\bibliographystyle{ieeetr}

\end{document}